\documentclass[superscriptaddress,groupedaddress,nofootnoteinbib,12pt]{article}  % for review and submission

\usepackage{amsmath}
\usepackage{amsfonts}
\usepackage{authblk}
\usepackage{bbm}
\usepackage{color}
\usepackage[dvipsnames]{xcolor}
\usepackage{graphicx,graphics}
\usepackage{epstopdf}
\usepackage{caption}
\usepackage{subcaption}
\usepackage{float} %This package is needed for the positioning the graphicx
\usepackage{pifont}
\usepackage{titlesec}
\usepackage{etoolbox}
\usepackage{indentfirst}
\usepackage{jcappub}
\usepackage{braket}
\usepackage{ulem}
\usepackage{dsfont}
\usepackage{slashed}
\usepackage{tikz}

\usepackage{bookmark,hyperref}
\hypersetup{
	colorlinks,
	citecolor=blue,
	filecolor=blue,
	linkcolor=blue,
	urlcolor=blue
}

\def\bb{\begin{equation}}
\def\ee{\end{equation}}
\def\ba{\begin{array}}
\def\ea{\end{array}}
\def\babc{\begin{subequations}}
\def\eabc{\end{subequations}}

\def\5{\hspace*{5mm}}
\def\2{{\scriptstyle\frac12}}

\tikzset{every node/.style={align=center}}

\begin{document}

%\begin{flushright}
%NYU-TH-14/03/2020
%\end{flushright}

\vspace{0.5in}

\begin{center}

\Large{\bf Torsion, Gravity Induced Chiral Symmetry Breaking and Cosmological Bounce}

\vspace{0.5in}

\large{Giorgi Tukhashvili}

\vspace{0.2in}

\large{\it Department of Physics, Princeton University, Princeton, NJ, 08544, USA}

\vspace{0.3in}

\end{center}

In this paper, we elaborate and extend our proposal \cite{Tukhashvili:2023itb} of the cosmological bouncing model in which a chiral condensate violates the null energy condition and induces a bounce. The condensate formation is caused by the gravitational effects that arise in the effective action when the curvature of space-time is non-zero. 
As the chiral condensation is a vacuum effect, the symmetries of the FRW space-time are not violated. We explicitly calculate the spinor effective action in the presence of a background gravitational field up to second order in curvature and with derivative corrections and also derive the initial conditions of the condensate. In addition, a new interaction term between the condensate and the scalar field is introduced that might be relevant for producing the adiabatic perturbations during the contraction phase.
The curvature/energy density stays far from Planckian values throughout this evolution. 

\vspace{3in}

\href{mailto:giorgit@princeton.edu}{giorgit@princeton.edu}

\newpage

%%%%%%%%%%%%%%%%%%%%%%%%%%%%%%%%%%%%%%%%%%%%%%%%%%%%%%%%%%%%%%%%%%%%%%%%%%
%%%%%%%%%%%%%%%%%%%%%%%%%%%%%%%%%%%%%%%%%%%%%%%%%%%%%%%%%%%%%%%%%%%%%%%%%%
%%%%%%%%%%%%%%%%%%%%%%%%%%%%%%%%%%%%%%%%%%%%%%%%%%%%%%%%%%%%%%%%%%%%%%%%%%

\section{\Large Introduction}

%%%%%%%%%%%%%%%%%%%%%%%%%%%%%%%%%%%%%%%%%%%%%%%%%%%%%%%%%%%%%%%%%%%%%%%%%%
%%%%%%%%%%%%%%%%%%%%%%%%%%%%%%%%%%%%%%%%%%%%%%%%%%%%%%%%%%%%%%%%%%%%%%%%%%
%%%%%%%%%%%%%%%%%%%%%%%%%%%%%%%%%%%%%%%%%%%%%%%%%%%%%%%%%%%%%%%%%%%%%%%%%%

Explaining the isotropy, homogeneity, and flatness of the Universe in a self-consistent way is the main challenge of modern cosmology. Recently, using numerical relativity, a novel method has been identified to solve these problems \cite{Cook:2020oaj,Ijjas:2020dws,Ijjas:2021gkf,Ijjas:2021wml,Kist:2022mew}: rather than invoke exponential expansion, utilize ultra-slow contraction of space. The matter content to achieve such dynamics consists of a canonical scalar with negative potential. This mechanism is robust, meaning that smoothness and flatness are achieved even if the initial state is far from smooth and flat, and it is rapid, meaning that the goal is achieved within a few $e$-folds \cite{Cook:2020oaj,Ijjas:2020dws,Ijjas:2021gkf,Ijjas:2021wml,Kist:2022mew}. The most remarkable feature of this mechanism is its simplicity, as it relies only on one large parameter, the largeness of which is a natural consequence of its definition.

From the observations, we have overwhelming evidence that the Universe is expanding today. If one invokes the ultra-slow contraction at the initial state to smooth and flatten the Universe, then to match the current observations, the space must stop the contraction and start to expand at some moment in time. That moment is called the cosmological bounce or simply the bounce. The singularity theorems by \textit{Penrose}, \textit{Hawking} and others \cite{Penrose:1964wq,Hawking:1966jv,Hawking:1970zqf} state that unless the matter driving the contraction violates the null energy condition (NEC), the space will crunch (the scale factor $a \rightarrow 0$) and singularities will develop. One way to avoid these is to argue that before the (big) crunch happens, new degrees of freedom associated with quantum gravity will emerge and prevent the development of singularities. Another way out is to introduce some form of matter to violate the NEC. The latter is the foundation for singularity theorems, and its violation might help avoid the crunching of space and as a result the quantum gravity regime. In this case, the scale factor, $a$, will decrease to some minimal value and then start to increase. This paper will refer to such an event as the classical bounce because the gravitational field stays classical during the evolution as the energies involved never hit the Planckian regime. 
\textit{Ijjas} and \textit{Steinhardt} \cite{Ijjas:2016tpn,Ijjas:2016vtq} showed that a classical and stable bounce could be achieved with a Galileon, a scalar field, the action of which involves higher derivative interactions; however, the equation of motion is still of the second order, i.e., no Ostrogradski instabilities. Earlier attempts to achieve a classical bounce using a galileon were made in \cite{Easson:2011zy,Cai:2012va,Cai:2013vm}, though these models experience instabilities \cite{Libanov:2016kfc} that are avoided in \cite{Ijjas:2016tpn,Ijjas:2016vtq}.

In this paper, we elaborate and extend the results of \cite{Tukhashvili:2023itb} and study the evolution of the Universe where a scalar field smoothes and flattens the Universe while the bounce is achieved using spinor condensate. So, the matter part of our model consists of a scalar field with negative exponential potential, a Dirac spinor, and a mixing term between these two. No on-shell particles are present, i.e., the chemical potential is zero. The latter is sufficient to avoid the inconsistencies in the previous works \cite{Trautman:1973wy,Isham:1974ci,Alexander:2008vt,Magueijo:2012ug} that include the direct violation of FRW symmetries and incompatibility of even a tiny spatial curvature with the physical fermions. The Dirac spinors source the torsion of space-time, which has to be accounted for by expanding ordinary Einsteins' General Relativity into Einstein-Cartan-Sciama-Kibble theory \cite{Sciama:1958we,Kibble:1961ba,Hehl:1976kj,Shapiro:2001rz}. In this formalism, torsion can be integrated out. Its remnants -- the four-Fermi interactions \cite{Weyl:1950xa} can be brought to Nambu-Jona-Lasinio (NJL) form \cite{Nambu:1961tp} by including non-minimal couplings between spinor currents and torsion \cite{Magueijo:2012ug,Farnsworth:2017wzr} and by properly choosing their coupling constants.
The coupling constant in front of the NJL potential, $\xi$, has tuned to be slightly smaller than its critical value (in flat space-time). We do not have justification for this tuning, which may be considered a downside of the model. Since the NJL coupling constant is less than its critical value, the chiral symmetry is not broken, spinor bi-linears have zero vev, and do not contribute to space-time evolution. The NJL condensation in our model is induced through the gravitational field. In Appendix \ref{app_loop_GR}, we calculate the effective action for the NJL model in the presence of a background gravitational field up to two orders in the curvature and with the derivative corrections. This generalizes the results of \cite{Inagaki:1993ya}, in which a similar calculation was done, but only to linear order in the curvature and without including the condensate derivatives. The calculation is done using both the Riemann normal coordinates and the Heat kernel method. Both results coincide in the UV limit. Our result shows that the negative values of the Ricci curvature can induce chiral symmetry breaking even if the NJL attraction is insufficient to form the condensates. 

The only assumptions we are making about the initial phase are that the space is contracting and the curvature is not large enough to induce the chiral symmetry breaking. During this phase, spinor bi-linears have zero vev and do not contribute to the energy budget. The dynamics is governed by the scalar field alone, which will take care of smoothing and flattening of the space \cite{Cook:2020oaj,Ijjas:2020dws,Ijjas:2021gkf,Ijjas:2021wml,Kist:2022mew}. After a few $e$-folds, the space will become homogeneous and isotropic, described by the spatially flat FRW metric. Using the scalar field to achieve smoothness and flatness is important because we can avoid having to rely on unjustified high-power spinor bi-linears to do this job \cite{Farnsworth:2017wzr}. The Ricci curvature is negative and its absolute value increases as space contracts. The curvature will eventually reach the critical value at which it causes chiral symmetry to break. The scalar bi-linear formed from the spinors acquire a non-zero vev, while the vevs of pseudo-scalar and vector bi-linears remain zero. Soon after this happens, the scalar field is no longer dominant, as within the energy budget, the contribution of the condensate becomes comparable to it. The bounce is what follows. Interestingly the sign of $\xi$ that causes attractive interaction between the spinors also guarantees the classicality of the bounce. 

Compared to \cite{Tukhashvili:2023itb}, the main differences in the model building are the potential for the scalar field, a new interaction that mixes the scalar and spinor bi-linear, and the absence of a rigid mass for the spinor. In \cite{Tukhashvili:2023itb}, we used a potential bounded from below. Here we use a negative exponential potential that transitions into the Morse potential after the chiral condensate formation due to the scalar-spinor mixing. All effective field theory (EFT) limits are respected throughout the evolution. 

This paper is organized as follows: in section \ref{sect_overview}, the action for our model is described, and the matter content and interactions are explained. The section \ref{Q_Average} is dedicated to discussing and demonstrating the curvature-induced NJL condensation. In section \ref{sect_Chron}, the timeline of the events is discussed, and the existence of the bounce is shown using the numerical solutions; in addition, the initial conditions for the fields are derived. The appendix \ref{app_int_tor} shows how to integrate out the torsion and obtain the NJL potential. Finally, the calculation of the NJL effective action in the presence of the gravitational field is presented in appendix \ref{app_loop_GR}.

Conventions through this paper are $(+,+,+)$ from Misner, Thorne and Wheeler \cite{Misner:1973prb}. Explicitly, the Minkowski metric is mostly plus, $\eta_{\mu \nu} = (-1,1,1,1)$, and the Riemann tensor is defined as ${R^\rho}_{\mu \lambda \nu} = + \partial_\lambda \Gamma^\rho_{\mu \nu} + \cdots$. The Planck constant will be assumed to have the unit value unless otherwise stated $\hbar =1$. The Planck mass is defined as $M_{pl} = \left( 8 \pi G \right)^{-1/2} \approx 2.5 \cdot 10^{18} GeV$, with $G$ being the gravitational Newtons constant. The Levi-Civita tensor $ \epsilon^{0 1 2 3} = 1 $. For the gamma matrices, the Weyl representation is used:
$\gamma^A = \left(
\begin{array}{cc}
	0 & \sigma^A \\
	\bar{\sigma}^A & 0 \\
\end{array}
\right) $ and $\gamma_5 = \left(
\begin{array}{cc}
	-1 & 0 \\
	0 & 1 \\
\end{array}
\right)$, with $\sigma^A = (1,\sigma^a)$ and $\bar{\sigma}^A =  (1,-\sigma^a)$. Here $\sigma^a$ are the Pauli matrices. The capital Latin letters represent the Local Lorentz indices having values $0,1,2,3$, while the lowercase Latin letters represent the Local Lorentz indices having values $1,2,3$.

\vspace{0.2in}

%%%%%%%%%%%%%%%%%%%%%%%%%%%%%%%%%%%%%%%%%%%%%%%%%%%%%%%%%%%%%%%%%%%%%%%%%%
%%%%%%%%%%%%%%%%%%%%%%%%%%%%%%%%%%%%%%%%%%%%%%%%%%%%%%%%%%%%%%%%%%%%%%%%%%
%%%%%%%%%%%%%%%%%%%%%%%%%%%%%%%%%%%%%%%%%%%%%%%%%%%%%%%%%%%%%%%%%%%%%%%%%%

\section{\Large Overview of Our Model}\label{sect_overview}

%%%%%%%%%%%%%%%%%%%%%%%%%%%%%%%%%%%%%%%%%%%%%%%%%%%%%%%%%%%%%%%%%%%%%%%%%%
%%%%%%%%%%%%%%%%%%%%%%%%%%%%%%%%%%%%%%%%%%%%%%%%%%%%%%%%%%%%%%%%%%%%%%%%%%
%%%%%%%%%%%%%%%%%%%%%%%%%%%%%%%%%%%%%%%%%%%%%%%%%%%%%%%%%%%%%%%%%%%%%%%%%%

This section defines the action our model is based on, and the interactions and matter content are justified. First, since the spinors are present, the first-order formalism of General Relativity (GR) must be used. Fundamental variables of the gravitational field are the one forms: the tetrad $e^A = e^A_\mu dx^\mu$ and the spin connection $\tilde{\omega}_{IJ}$. The metric tensor is defined using the tetrads: $g_{\mu \nu } = \eta_{AB} e_\mu^A e_\nu^B$, where $\eta_{AB}$ is the Minkowski metric. The spinor sector consists of a Dirac spinor with one color/flavor. Then, to justify using the FRW metric during the bounce phase, we need a field during the initial phase of contraction that enables smoothing and flattening. We entrust this mission to the scalar field with a negative exponential potential, as commonly used in models of ultra-slow contraction \cite{Cook:2020oaj,Ijjas:2020dws,Ijjas:2021gkf,Ijjas:2021wml}. Lastly, there should exist a mixing between the scalar and spinor. The latter should be responsible for the scalar to condensate energy transfer and, ideally, causing an effective friction for the scalar field, which is required to produce the adiabatic perturbations during the contraction \cite{Ijjas:2021ewd,Ijjas:2020cyh}. Once the condensate is formed, the mixing term also contributes to the potential of the scalar and bounds it from below (See Fig. \ref{exp_to_morse}). Our action:
\begingroup\makeatletter\def\f@size{11}\check@mathfonts
\begin{align}\label{orig_action}
	\nonumber  \mathcal{S} = & - \frac{M_{pl}^2}{4} \epsilon_{A B M N}
	e^A \wedge e^B \wedge \tilde{R}^{M N} - \frac{i}{2 \cdot 3!} \epsilon_{A B M N} e^A \wedge e^B \wedge e^M \wedge  \Big( \overline{\Psi} \gamma^N \tilde{D} \Psi - \tilde{D} \overline{\Psi} \gamma^N \Psi \Big)  \\
	{} &	+ \sqrt{\frac{\xi}{6}} \epsilon_{I J K L} \tilde{D} e^I \wedge e^J \wedge e^K ~V^L
	+ \left( \frac{1}{4} - \sqrt{\frac{\xi}{6}} \right) \tilde{D} e^I \wedge e_I \wedge e_J ~A^J  \\
	\nonumber {} & + \frac{1}{4!}  \epsilon_{A B M N} e^A \wedge e^B \wedge e^M \wedge e^N
	\left[ \frac12 (\partial \phi)^2 + V (\phi) + F ( \phi ) \overline{\Psi} \Psi \right] .
\end{align}
\endgroup
The integration sign is implicit in front of each term. The quantities with a tilde are introduced to distinguish them from torsion-free ones (without a tilde). The first term is the torsion-modified Einstein-Hilbert, and the second is the action for the massless Dirac spinor in the presence of the gravitational field. The Dirac field is assumed to have one color and one flavor, but generalization to $N$ colors/flavors is straightforward. The third and fourth terms are the non-minimal couplings between the torsion and spinor currents. The third line describes a scalar field with the potential $V (\phi) $ coupled minimally to gravity. The very last term is the coupling between the spinor and the scalar. $F (\phi)$ is not necessarily a linear function, i.e., Yukawa. The coupling constants on the second line are chosen so that, after the torsion is integrated out, they induce chiral symmetry. This choice is not unique. The covariant derivatives and the currents are defined as:
\bb
\tilde{D} \Psi = \left( d - \frac{1}{8} \tilde{\omega}_{AB} \left[ \gamma^A , \gamma^B \right] \right) \Psi ;
\5\5\5
\tilde{D} \overline{\Psi} = \overline{\Psi} \left( \overleftarrow{d} + \frac{1}{8} \tilde{\omega}_{AB} \left[ \gamma^A , \gamma^B \right] \right) ;
\ee
\bb
\tilde{D} e^I = d e^I + \left. \tilde{\omega}^I \right.{}_J \wedge e^J = T^I .
\ee
\bb
V^M \equiv \overline{\Psi} \gamma^M \Psi ;
\5\5\5
A^M \equiv \overline{\Psi} \gamma_5 \gamma^M \Psi .
\ee
Since, in the absence of a source, the torsion, $T^I$, vanishes and we recover the ordinary Einstein-Hilbert action, it is possible to postulate that the connection $\tilde{\omega}_{IJ}$ consists of two parts: $\tilde{\omega}_{IJ} = \omega_{IJ} + C_{IJ}$. The first part is the torsion-free connection $\omega_{IJ}$, while the second part is the tensor, $C_{IJ}$, related to the torsion as $T^I = \left. C^I \right.{}_J \wedge e^J$. From now on, the quantities without tilde will be defined wrt the torsion-free connection $\omega_{IJ}$. Since, by definition, $T^I$ and $C_{IJ}$ are linearly related, we will refer to both as torsion. Once this philosophy is adopted, if the gravitation (curvature) is turned off, (\ref{orig_action}) will not just reduce to the action of a spinor field coupled to a scalar. Additional interactions arise because torsion stays non-zero even when the curvature vanishes. The tensor $C_{IJ}$ enters algebraically into the action (\ref{orig_action}); in addition, its highest degree is quadratic. This allows us to integrate it out from (\ref{orig_action}). The equivalent action takes the following form (see Appendix \ref{app_int_tor}):
\begingroup\makeatletter\def\f@size{11}\check@mathfonts
\begin{align}\label{action_torsion_IO}
	\nonumber  \mathcal{S} = & - \frac{M_{pl}^2}{4} \epsilon_{A B M N}
	e^A \wedge e^B \wedge R^{M N} - \frac{i}{2 \cdot 3!} \epsilon_{A B M N} e^A \wedge e^B \wedge e^M \wedge \Big( \overline{\Psi} \gamma^N D \Psi - D \overline{\Psi} \gamma^N \Psi \Big)  \\
	{} &	- \frac{\xi}{4! M_{pl}^2}  \epsilon_{A B M N} e^A \wedge e^B \wedge e^M \wedge e^N
	\left[ \big( \overline{\Psi} \Psi \big)^2 + \big( i \overline{\Psi} \gamma_5 \Psi \big)^2 \right]  \\
	\nonumber {} & + \frac{1}{4!}  \epsilon_{A B M N} e^A \wedge e^B \wedge e^M \wedge e^N
	\left[ \frac12 (\partial \phi)^2 + V (\phi) + F ( \phi ) \overline{\Psi} \Psi \right] .
\end{align}
\endgroup
The action (\ref{action_torsion_IO}) is fully equivalent to (\ref{orig_action}). Torsion was exchanged for the four-Fermi interactions. The specific form of these interactions is a consequence of the choice of coupling constants on the second line of (\ref{orig_action}) and the Fierz identity:
\bb\label{Fierz_id}
V_K V^K - A_K A^K = 2 \left[ \big( \overline{\Psi} \Psi \big)^2 + \big( i \overline{\Psi} \gamma_5 \Psi \big)^2 \right] .
\ee
Ignoring the mixing with scalar, the spinor content of (\ref{action_torsion_IO}) is equivalent to the NJL model. The Einstein and scalar equations are obtained by taking the variation of (\ref{action_torsion_IO}) wrt $e^A$ and $\phi$ respectively (diffeomorphism and Local Lorentz indices can be mapped between each other by using $e^A_\mu; ~ e_A^\mu$):
\begin{align}\label{einstein}
	\nonumber {} & - M_p^2 G^{X}_A + i \delta^X_A \overline{\Psi} \gamma^N D_N \Psi
	- i \overline{\Psi} \gamma^X D_A \Psi + \delta^X_A \frac{\xi}{ M_{pl}^2} \left[ \big( \overline{\Psi} \Psi \big)^2 + \big( i \overline{\Psi} \gamma_5 \Psi \big)^2 \right]  \\
	{} & + \partial_A \phi \partial^X \phi - \delta_A^X \Big( \frac12 (\partial \phi)^2 + V (\phi) + F (\phi) \overline{\Psi} \Psi \Big) = 0 ;
\end{align}
\bb\label{scalar_eq_1}
- \nabla^2 \phi + \frac{d V}{d \phi} + \frac{d F}{d \phi} \overline{\Psi} \Psi = 0 .
\ee

In the simplest setup of Einstein-Cartan-Sciama-Kibble theory, if the Dirac spinor sources the torsion, the spin-spin interaction is $A_\mu A^\mu$ \cite{Kibble:1961ba,Weyl:1950xa}, where $A_\mu$ is the axial current of the spinor. \textit{Trautman} \cite{Trautman:1973wy} noticed that if the axial current vector is time-like, then these interactions can source NEC violation, and the cosmic singularity can be avoided.

The phenomenological implications of torsion were studied in \cite{Bagrov:1994re,Hammond:1995rp,Hammond:1996ua}. It turns out that its presence induces a potential similar to the electromagnetic field. Within the hydrogen atom, torsion interacting with the electron's spin will cause the splitting of spectral lines, similar to the Zeeman effect. In \cite{Hammond:1995rp}, this fact was used to constrain the coupling constant related to the torsion. In our notation, their constraint roughly translates to $\sqrt{\xi} < 10^{17}$. A more stringent constraint can be achieved by considering the interaction between polarized test mass and paramagnetic salt \cite{Chui:1993jp,Hammond:1995rp}, in this case, $\sqrt{\xi} < 10^{14}$. In the numerical calculations done in section \ref{numerics}, $\xi \approx 200$, which easily satisfies these constraints.

%%%%%%%%%%%%%%%%%%%%%%%%%%%%%%%%%%%%%%%%%%%%%%%%%%%%%%%%%%%%%%%%%%%%%%%%%%
%%%%%%%%%%%%%%%%%%%%%%%%%%%%%%%%%%%%%%%%%%%%%%%%%%%%%%%%%%%%%%%%%%%%%%%%%%
%%%%%%%%%%%%%%%%%%%%%%%%%%%%%%%%%%%%%%%%%%%%%%%%%%%%%%%%%%%%%%%%%%%%%%%%%%

\section{\Large Curvature Induced Spinor Condensation}\label{Q_Average}

%%%%%%%%%%%%%%%%%%%%%%%%%%%%%%%%%%%%%%%%%%%%%%%%%%%%%%%%%%%%%%%%%%%%%%%%%%
%%%%%%%%%%%%%%%%%%%%%%%%%%%%%%%%%%%%%%%%%%%%%%%%%%%%%%%%%%%%%%%%%%%%%%%%%%
%%%%%%%%%%%%%%%%%%%%%%%%%%%%%%%%%%%%%%%%%%%%%%%%%%%%%%%%%%%%%%%%%%%%%%%%%%

Through this paper, the gravitational field is treated classically. To have a self-consistent description, the spinor bi-linears that interact with gravity must be averaged wrt the vacuum state. The average of these bi-linears is non-zero if the spinor number density is itself non-zero, but in this case, unless these spinors are arranged in a specific way, the symmetries of the FRW space will not be respected. To get around this problem, the spinor number density will be assumed to vanish $\braket{\overline{\Psi} \gamma^0 \Psi}=0$. Instead, since the four-Fermi interactions are present, the NJL mechanism is invoked to justify the non-zero VEV (quantum average) of $\braket{\overline{\Psi} \Psi}$. 
We will assume $\braket{ \left( \overline{\Psi} \Psi \right) ^2} = \braket{ \overline{\Psi} \Psi }^2$ to be a good approximation, even though the equality strictly holds in the large $N $ limit, with difference scaling as $N^{-2}$ \cite{Shifman:1978bx}.

Since the background space-time is not flat, the corrections from the non-zero curvature within the effective action need to be accounted for. Appropriate calculation of the effective action is done in the Appendix \ref{app_loop_GR}, where the Riemann normal coordinates are used to show that the effective action up to two orders in the Riemann curvature is given by the formula (\ref{eff_action_CC_kin}). Even though these corrections have a quantum nature (proportional to $\hbar$), Planck's constant arises from the spinor loop, while the gravitational field is classical. Also, these new terms should not be understood as corrections to Einstein-Hilbert. The curvatures in (\ref{eff_action_CC_kin}) act as a background field that can cause or prevent condensate formation (the latter can be compared to the Cooper pairs in an external magnetic field). The gap equation with the curvature corrections:
\begin{align}\label{gap_eq_orig}
	\nonumber {} & - \frac{M_{pl}^2}{\Lambda^2} \frac{8 \pi^2}{\xi} u + \frac{\partial f_4 }{\partial u}  - \frac{2}{\Lambda^2} f_0 ~\nabla^2 u - \frac{1}{\Lambda^2} \left( \partial u \right)^2 \frac{\partial f_0 }{\partial u}
	+ \frac{R }{6 \Lambda^2} \left( 2 u f_1 + u^2 \frac{\partial f_1 }{\partial u} \right)   \\
	\nonumber {} & + \frac{2}{\Lambda^4} \left( - \frac{1}{120} \nabla^2 R + \frac{1}{288} R^2
	- \frac{1}{180} R_{\mu \nu} R^{\mu \nu}
	- \frac{7}{1440} R_{\mu \nu \rho \sigma} R^{\mu \nu \rho \sigma} \right) \frac{\partial f_1 }{\partial u}  \\
	{} & - \frac{2}{\Lambda^4} \left( \frac{1}{12} \nabla^2 R + \frac{1}{36} R_{\mu \nu} R^{\mu \nu}
	+ \frac{1}{144} R_{\mu \nu \rho \sigma} R^{\mu \nu \rho \sigma} \right) \frac{\partial f_2 }{\partial u}  \\
	\nonumber {} & + \frac{2}{\Lambda^4} \left( \frac{1}{10} \nabla^2 R + \frac{1}{72} R^2
	+ \frac{7}{180} R_{\mu \nu} R^{\mu \nu} + \frac{1}{60} R_{\mu \nu \rho \sigma} R^{\mu \nu \rho \sigma} \right) \frac{\partial f_3 }{\partial u} = 0 .
\end{align}
The functions $f_i$ and the dimensionless variables are defined in (\ref{aux_field_def}), (\ref{def_f_1_2}-\ref{def_dimless_vars}) and (\ref{def_f_0}). The derivative terms are necessary as it is no longer valid in a dynamic space-time to assume the condensate to be constant. Note that this equation and (\ref{eff_action_CC_kin}) involve a new parameter $y \equiv \epsilon / \Lambda$, where $\epsilon$ is the infrared cut-off. The necessity of IR cut-off can be seen from the fact that the effective action
\bb\label{def_eff_ac_log}
\mathcal{S}_{eff} \propto \log \frac{\det{\left( i \slashed{D} - \sigma \right)} }{\det{\left( i \slashed{D} \right)}}
\ee
involves the loop of a massless spinor. In the zeroth and first order curvature part of the (\ref{eff_action_CC_kin}), the IR scale plays no role and can be set to zero, while in the part proportional to the curvature squared, the IR scale is necessary for the effective action to be regular at $\sigma=0$, see (\ref{def_eff_ac_log}). 
The necessity of the IR cut-off disappears if one accounts for the effect of explicit chiral symmetry breaking already present within (\ref{action_torsion_IO}). The explicit symmetry breaking was ignored while calculating the effective action (\ref{eff_action_CC_kin}). As the role $\epsilon$ is to partially replace $F (\phi)$ in (\ref{eff_action_CC_kin}), it is fair to assume these two to be of the same order of magnitude at the time of spontaneous chiral symmetry breaking.

\begin{figure}[t]
	\centering
	\begin{subfigure}[b]{0.49\textwidth}
	\includegraphics[width=\textwidth]{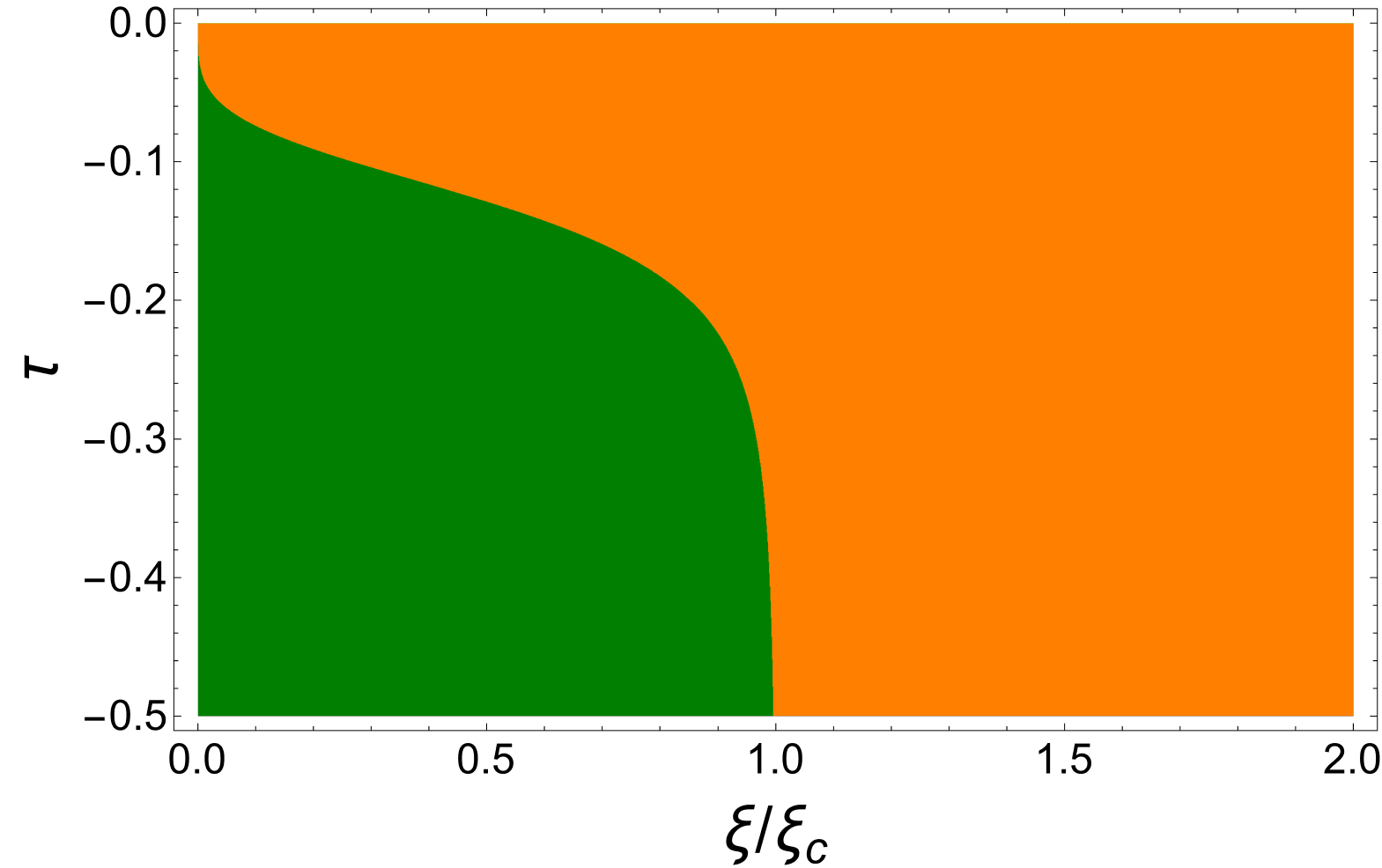}
	\end{subfigure}
	\begin{subfigure}[b]{0.49\textwidth}
	\includegraphics[width=\textwidth]{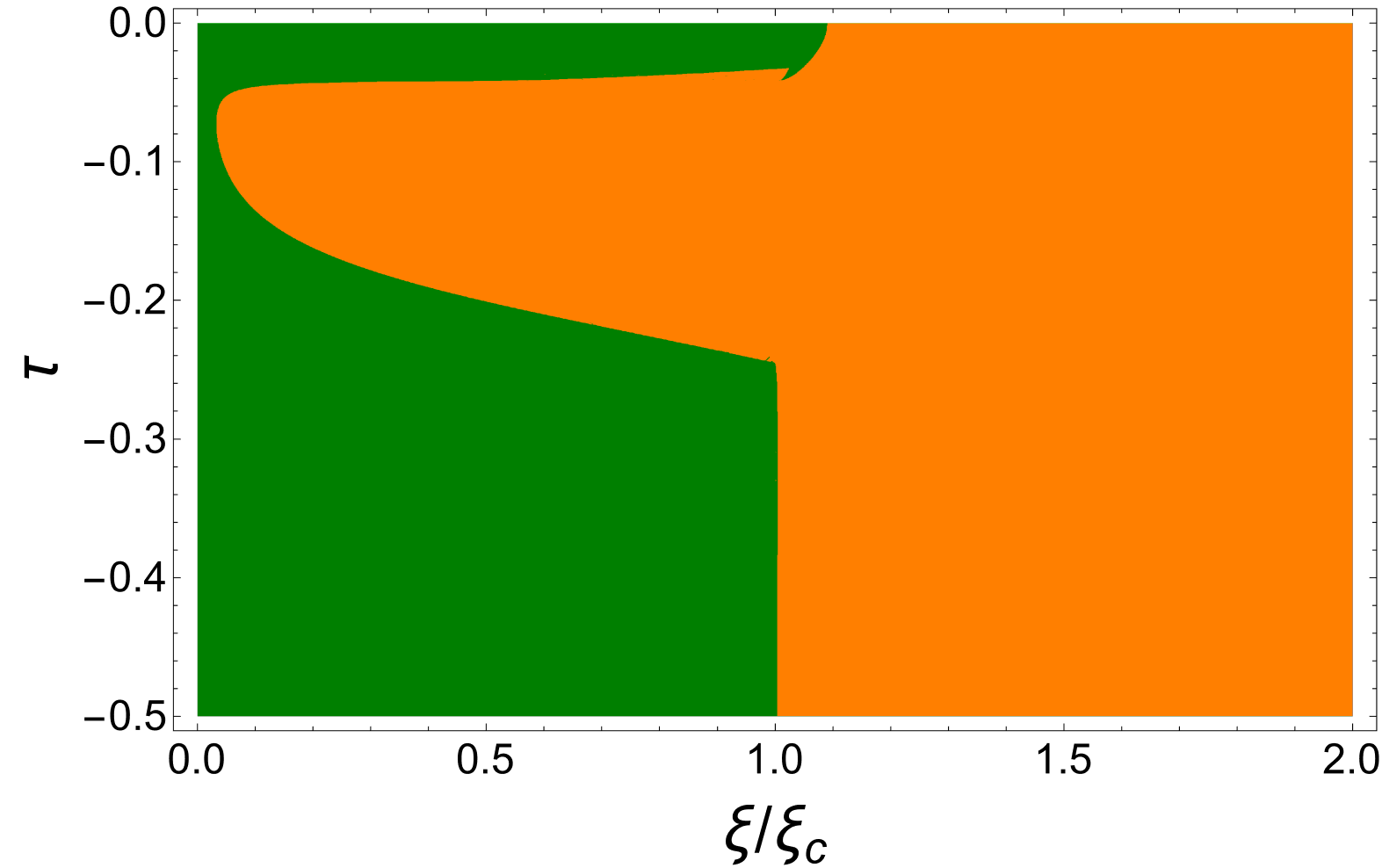}
	\end{subfigure}
   \caption{Phase diagrams of the NJL model in an external gravitational field according to (\ref{eff_action_CC}). Orange, the variation of (\ref{eff_action_CC}) has a non-trivial solution meaning that the chiral symmetry is broken. Green, the chiral symmetry is unbroken. If the curvature corrections are ignored, the two phases will be separated with a vertical line $\xi / \xi_c =1$. In both cases, the space-time is spatially flat FRW. Left: the Hubble parameter $ = 2 / (\kappa^2 \tau)$ with $\kappa = 10$, $y=10^{-3}$ and $\theta = 10^{-3}$; see (\ref{var_defs}) and (\ref{par_defs}). Right: the Hubble parameter has Trautman ansatz, $2 \tau / \kappa^2 (10^{-2} + \tau^2)$. In both cases, the back-reaction of the condensate to space-time and the limits of the curvature expansion in (\ref{eff_action_CC_kin}) are ignored.}\label{phase_diags}
\end{figure}

In the absence of gravity, condensate formation is determined by the ratio $\xi_c /\xi$ where $\xi_c \equiv 2 \pi^2 M_{pl}^2 / \Lambda^2 $ is the critical value of the four-Fermi interactions when space-time is flat and $y=0$. For $\xi_c /\xi < 1 $ the chiral symmetry is spontaneously broken. Once gravity is introduced into the picture, the story might change. If this ratio is tuned to be near one, then in the presence of positive Ricci curvature, the chiral symmetry might be restored even if $\xi_c /\xi < 1 $, while in the presence of negative Ricci curvature, the chiral symmetry can be broken with $\xi_c /\xi > 1 $. 
The value of Ricci curvature at which the chiral symmetry breaking happens is referred as the critical curvature, $R_c$. The critical curvature depends on the value of NJL coupling, $R_c (\xi)$. If this coupling is taken to be its critical value (in flat space-time), then even a tiny negative curvature can induce the chiral symmetry breaking, $R_c (\xi_c) = 0$.
Here, we have not taken into account the running of $\xi$, which itself increases with increasing curvature \cite{Shapiro:1994vs}. This effect should enhance the process of condensation.
Figure \ref{phase_diags} shows the phase diagrams when the space-time is spatially flat FRW for different Hubble parameters.

The dynamical symmetry breaking due to the curvature contributions within the radiative corrections in the massless scalar model with quartic self-interactions was studied in \cite{Ford:1981xj}.
The curvature-induced NJL condensation should not be confused with BCS condensation by \textit{Alexander} and \textit{Biswas} \cite{Alexander:2008vt} (see also \cite{Alexander:2014uaa}). Similar to \cite{Trautman:1973wy}, in \cite{Alexander:2008vt}, there is an initial non-zero density of physical particles. In \cite{Alexander:2008vt}, the condensation catalyst is the shrinking volume, while in our case -- the increasing curvature. 

%%%%%%%%%%%%%%%%%%%%%%%%%%%%%%%%%%%%%%%%%%%%%%%%%%%%%%%%%%%%%%%%%%%%%%%%%%
%%%%%%%%%%%%%%%%%%%%%%%%%%%%%%%%%%%%%%%%%%%%%%%%%%%%%%%%%%%%%%%%%%%%%%%%%%
%%%%%%%%%%%%%%%%%%%%%%%%%%%%%%%%%%%%%%%%%%%%%%%%%%%%%%%%%%%%%%%%%%%%%%%%%%

\section{\Large Chiral Condensate Mediated Cosmological Bounce}\label{sect_Chron}

In the last section, we showed that the negative Ricci curvature could induce chiral symmetry breaking in the NJL model even when the four-Fermi coupling strength is insufficient to form the condensate. In this section, this fact is used, together with the assumption that the four-Fermi coupling is slightly less than its critical value, such that, when the curvature is small, the chiral symmetry is not broken and $\braket{\overline{\Psi} \Psi} = 0$. \\

\begin{figure}[b]
	\centering
	\begin{subfigure}[b]{0.49\textwidth}
	\includegraphics[width=\textwidth]{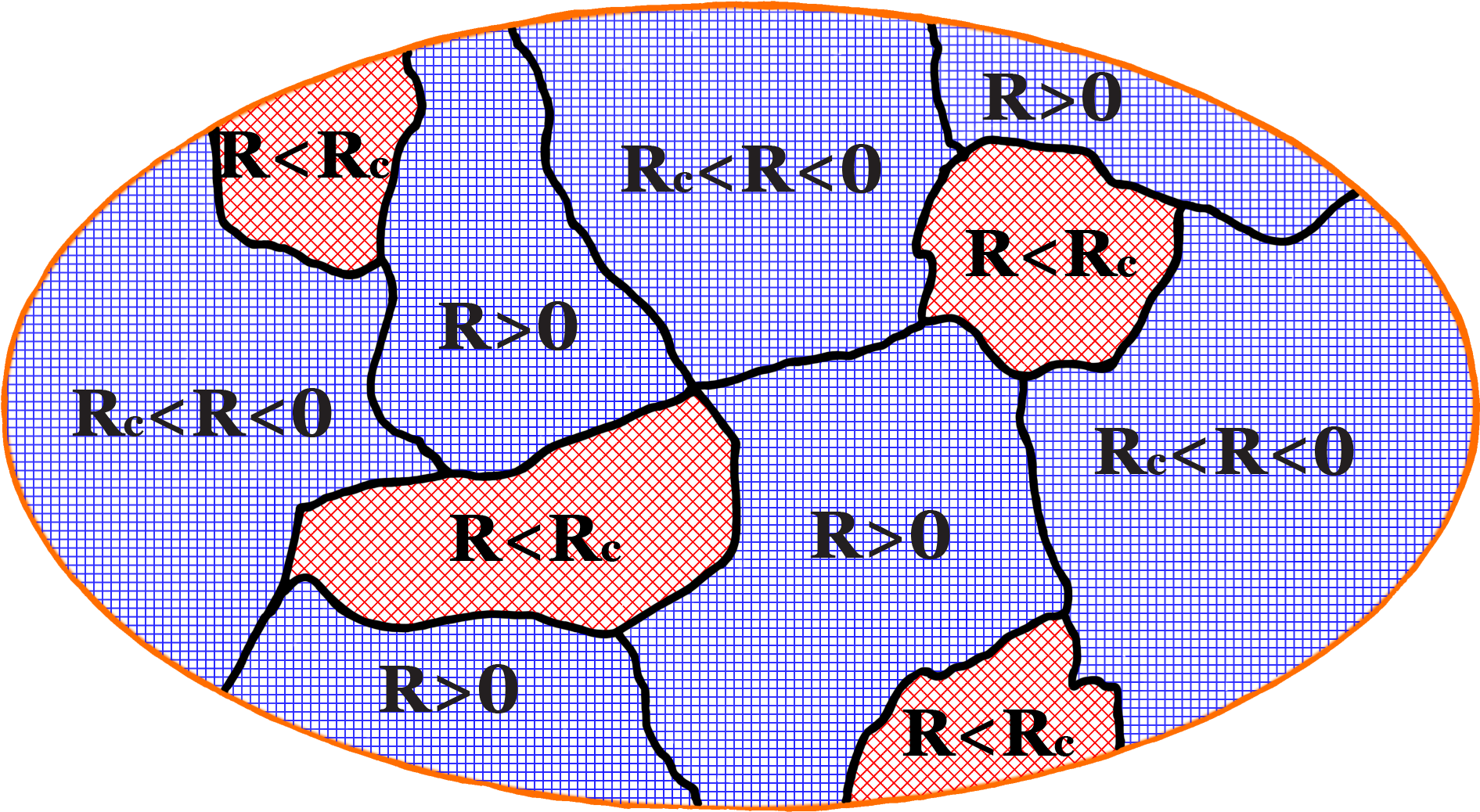}
    \caption{}\label{uni_sketch_left}
	\end{subfigure}
	\begin{subfigure}[b]{0.49\textwidth}
	\includegraphics[width=\textwidth]{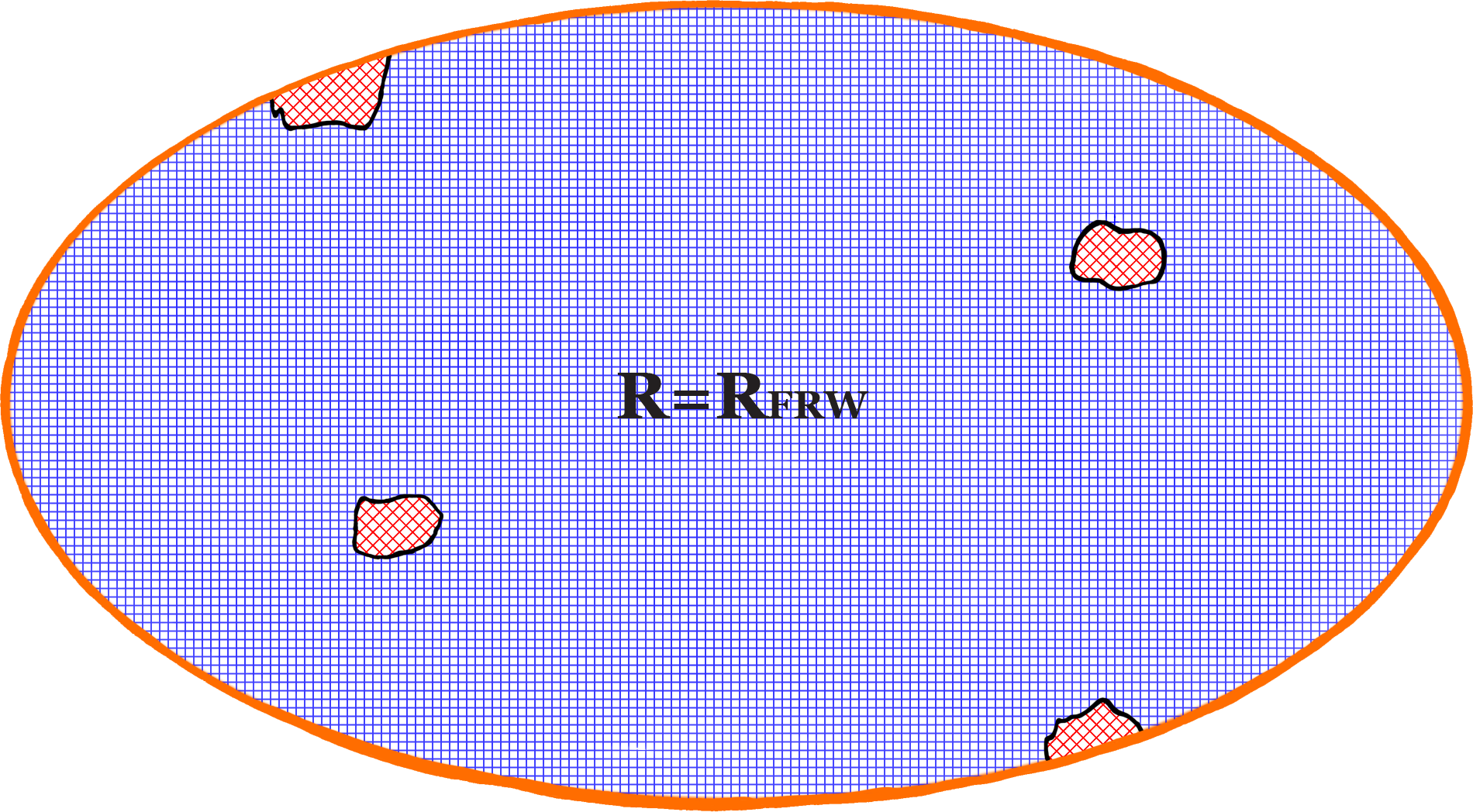}
    \caption{}\label{uni_sketch_right}
	\end{subfigure}
   \caption{Left: Initial state of the universe is partitioned into different patches with approximately the same curvature. Right: The universe after undergoing the super-smoothing mechanism \cite{Cook:2020oaj,Ijjas:2020dws,Ijjas:2021gkf,Ijjas:2021wml,Kist:2022mew}. Most of the space, depicted by the blue + pattern, can be well approximated by the flat FRW metric.}\label{uni_sketch}
\end{figure}

At this point, we are ready to present the cosmological bounce scenario. Following \cite{Tukhashvili:2023itb}, the only assumption about the initial state is that space is contracting; otherwise, the initial conditions are inhomogeneous and anisotropic, non-perturbatively far from FRW. The universe can be partitioned into different patches with approximately the same Ricci curvature; see Fig. \ref{uni_sketch_left}. If the Ricci curvature of the patch is positive $(R>0)$, or negative, but larger than the critical curvature $(0>R>R_c (\xi))$, then within this patch the chiral symmetry is not broken, the spinor bi-linears have zero vev and the scalar field entirely defines the dynamics. These regions are colored with the blue $+$ pattern in Fig. \ref{uni_sketch}. If, on the other hand, the Ricci curvature of the patch is less than the critical value $(R<R_c (\xi))$, then within this patch, the chiral symmetry is broken, and the dynamics are defined by both, the scalar and the spinor fields. These regions are colored with the red $\times$ pattern in Fig. \ref{uni_sketch}.
Within the blue regions, the scalar field controls the dynamics; and by using the results of \cite{Cook:2020oaj,Ijjas:2020dws,Ijjas:2021gkf,Ijjas:2021wml,Kist:2022mew}, these regions will rapidly converge to FRW. Once this happens these regions shrink very slowly $a \propto (t^2)^{1/(2 \epsilon_\phi)}$, as $\epsilon_\phi \gg 1$. The dynamics of the red regions are completely different, and to get the complete picture, one needs to perform numerical studies similar to \cite{Cook:2020oaj,Ijjas:2020dws,Ijjas:2021gkf,Ijjas:2021wml,Kist:2022mew}. The latter goes beyond the scope of this paper, but to get a glimpse of what might happen, let's assume that the space is flat, homogeneous, and anisotropic. Here are three possible scenarios:

1) The equation of state of the dominant matter $\epsilon < 3$; in this case, the mixmaster behavior will develop, and anisotropy will dominate. Parametrizing the anisotropies with the Kasner-like metric $g_{\mu \nu} = (-1, a^2 e^{2 \beta_1}, a^2 e^{2 \beta_2}, a^2 e^{2 \beta_3})$, the Ricci scalar gets a new contribution:
\bb
R = 6 \left( \frac{\dot{a}^2}{a^2} + \frac{\ddot{a}}{a}  \right)
+ \dot{\beta}_1^2 + \dot{\beta}_2^2 + \dot{\beta}_3^2 .
\ee
This formula shows that the anisotropy contributes positively, and the growth of this term favors the restoration of the chiral symmetry. Once this happens, the red region becomes light blue; we have already discussed what occurs next.

2) The equation of state of the dominant matter $\epsilon \gg 3$. Even though the results of \cite{Cook:2020oaj,Ijjas:2020dws,Ijjas:2021gkf,Ijjas:2021wml,Kist:2022mew} are based on a scalar field with a large equation of state, the smoothing is likely to happen with any type of matter with a large equation of state \cite{Erickson:2003zm}, so the same outcome $\rightarrow$ spatially flat FRW after a few $e$-folds is expected.

3) The equation of state of the dominant matter $\epsilon \gtrsim 3$, this case is more complicated, and even the assumptions we made about the initial state are insufficient to make any predictions.

In any case, since the equation of state of the dominant matter is $\epsilon \approx 3$, red regions shrink faster than blue regions. After a while, Fig. \ref{uni_sketch_left} evolves into Fig. \ref{uni_sketch_right}, where the blue regions dominate the space. 

The remaining paper will focus solely on the blue regions in Figure \ref{uni_sketch_right}, which are homogeneous and isotropic (FRW) to a good approximation. Within these regions, the following system of equations determines the dynamics of the fields $(H \equiv \dot{a}/a)$:
\bb\label{scalar_in_FRW}
\ddot{\phi} + 3 \frac{\dot{a}}{a} \dot{\phi} + \frac{d V}{d \phi} + \frac{d F}{d \phi} \braket{\overline{\Psi} \Psi} = 0 ;
\ee
\bb
3 M_{pl}^2 H^2 = - \frac{\xi}{M_p^2} \braket{\overline{\Psi} \Psi}^2 + \frac12 \dot{\phi}^2 + V (\phi) + F(\phi) \braket{\overline{\Psi} \Psi} ;
\ee
\begin{align}\label{condensate_in_FRW}
	\nonumber {} & - \frac{4 \xi_c}{\xi} u + \frac{\partial f_4 }{\partial u}  + \frac{2}{\Lambda^2} f_0 \left( \ddot{u} + 3 H \dot{u} \right) + \frac{1}{\Lambda^2} \dot{u}^2 \frac{\partial f_0 }{\partial u}
	+ \frac{1}{\Lambda^2} \left( 2 H^2 + \dot{H} \right) \left( 2 u f_1 + u^2 \frac{\partial f_1 }{\partial u} \right)   \\
	\nonumber {} & + \frac{2}{\Lambda^4} \left( \frac{1}{20} \dddot{H} + \frac{7}{20} H \ddot{H} + \frac{47}{60} H^2 \dot{H} + \frac{1}{5} \dot{H}^2 + \frac{11}{60} H^4 \right) \frac{\partial f_1 }{\partial u}  \\
	{} & - \frac{2}{\Lambda^4} \left( - \frac{1}{2} \dddot{H} - \frac{7}{2} H \ddot{H} - \frac{29}{6} H^2 \dot{H} - \frac{19}{12} \dot{H}^2 + \frac{7}{6} H^4 \right) \frac{\partial f_2 }{\partial u}  \\
	\nonumber {} & + \frac{2}{\Lambda^4} \left( - \frac{3}{5} \dddot{H} - \frac{21}{5} H \ddot{H} - \frac{17}{5} H^2 \dot{H} - \frac{37}{30} \dot{H}^2 + \frac{19}{5} H^4 \right) \frac{\partial f_3 }{\partial u} = 0 .
\end{align}
The latter is the gap equation (\ref{gap_eq_orig}) in the FRW space-time and $u = - 2 \xi \braket{\overline{\Psi} \Psi}/ (\Lambda M_{pl}^2 )$.

%%%%%%%%%%%%%%%%%%%%%%%%%%%%%%%%%%%%%%%%%%%%%%%%%%%%%%%%%%%%%%%%%%%%%%%%%%
%%%%%%%%%%%%%%%%%%%%%%%%%%%%%%%%%%%%%%%%%%%%%%%%%%%%%%%%%%%%%%%%%%%%%%%%%%
%%%%%%%%%%%%%%%%%%%%%%%%%%%%%%%%%%%%%%%%%%%%%%%%%%%%%%%%%%%%%%%%%%%%%%%%%%

\subsection{\large Numerical Solutions}\label{numerics}

%%%%%%%%%%%%%%%%%%%%%%%%%%%%%%%%%%%%%%%%%%%%%%%%%%%%%%%%%%%%%%%%%%%%%%%%%%
%%%%%%%%%%%%%%%%%%%%%%%%%%%%%%%%%%%%%%%%%%%%%%%%%%%%%%%%%%%%%%%%%%%%%%%%%%
%%%%%%%%%%%%%%%%%%%%%%%%%%%%%%%%%%%%%%%%%%%%%%%%%%%%%%%%%%%%%%%%%%%%%%%%%%

The potential for the scalar field is a negative exponential, which allows to achieve a smooth and flat space. The ansatz of the mixing function, $F$, is also chosen to be an exponent.
\bb
V (\phi) = - M_{pl}^2 V_0^2 \exp \left( - \frac{\kappa}{M_{pl}} \phi \right);
\5\5\5
F(\phi) =  F_0 \exp \left( - n \frac{\kappa}{M_{pl}} \phi \right).
\ee
The parameter $\kappa \equiv M_{pl} / \Lambda_\phi$ controls the number of $e$-folds of contraction of the Hubble radius required to achieve the smooth and flat state. 
The large value of $\kappa$ is a consequence of its definition, and increasing this value enhances both rapidity and robustness.
$F_0$ has the dimension of mass, and $n \geq 1$. To solve the equations (\ref{scalar_in_FRW})-(\ref{condensate_in_FRW}) numerically, it is convenient to define dimensionless variables and parameters:
\bb\label{var_defs}
\varphi \equiv \frac{\phi}{M_{pl}};
\5\5\5
h \equiv \frac{a'}{a} ;
\5\5\5
\tau \equiv V_0 ~t ;
\5\5\5
{}' \equiv \frac{d}{d \tau} ;
\ee
\bb\label{par_defs}
\theta \equiv \frac{V_0^2}{\Lambda^2} ;
\5\5\5
\bar{F}_0 \equiv \frac{ F_0 }{2 \xi \theta \Lambda} ;
\5\5\5
\xi_c \equiv 2 \pi^2 \frac{M_{pl}^2}{\Lambda^2} .
\ee
In terms of the new variables, the system (\ref{scalar_in_FRW})-(\ref{condensate_in_FRW}) takes the following shape:
\bb\label{scalar_num_form}
\varphi'' + 3 h \varphi' + \kappa e^{- \kappa \varphi} + n \kappa \bar{F}_0 e^{- n \kappa \varphi} u =0 ;
\ee
\bb\label{Friedmann_num_form}
3 h^2 =  - \frac{1}{4 \xi \theta} u^2
+ \frac12 \left( \varphi ' \right)^2 - e^{- \kappa \varphi}
- \bar{F}_0 e^{- n \kappa \varphi} u ;
\ee
\begin{align}\label{condensate_num_form}
	\nonumber {} & u'' + 3 h u' + \left( u' \right)^2 \frac{1}{2 f_0} \frac{\partial f_0}{\partial u}
	+ \frac{1}{2 \theta f_0} \frac{\partial f_4}{\partial u} - \frac{2 \xi_c}{\xi \theta f_0}  u
	+ \frac{1}{2} \left( 2 h^2 + h' \right) \left( 2 u \frac{f_1}{f_0} + \frac{u^2}{f_0} \frac{\partial f_1}{\partial u} \right)  \\
	{} & + \theta \left( \frac{1}{20} h''' + \frac{7}{20} h h'' + \frac{47}{60} h^2 h' + \frac{1}{5} (h')^2 + \frac{11}{60} h^4 \right) \frac{1}{f_0} \frac{\partial f_1}{\partial u}
	 \\
	\nonumber {} & - \theta \left( - \frac{1}{2} h''' - \frac{7}{2} h h'' - \frac{29}{6} h^2 h' - \frac{19}{12} (h')^2 + \frac{7}{6} h^4 \right) \frac{1}{f_0} \frac{\partial f_2}{\partial u}  \\
	\nonumber {} & + \theta \left( - \frac{3}{5} h''' - \frac{21}{5} h h'' - \frac{17}{5} h^2 h' - \frac{37}{30} (h')^2 + \frac{19}{5} h^4 \right)
	\frac{1}{f_0} \frac{\partial f_3}{\partial u} = 0 .
\end{align}

\begin{figure}[t]
	\centering
	\begin{subfigure}[b]{0.49\textwidth}
		\includegraphics[width=\textwidth]{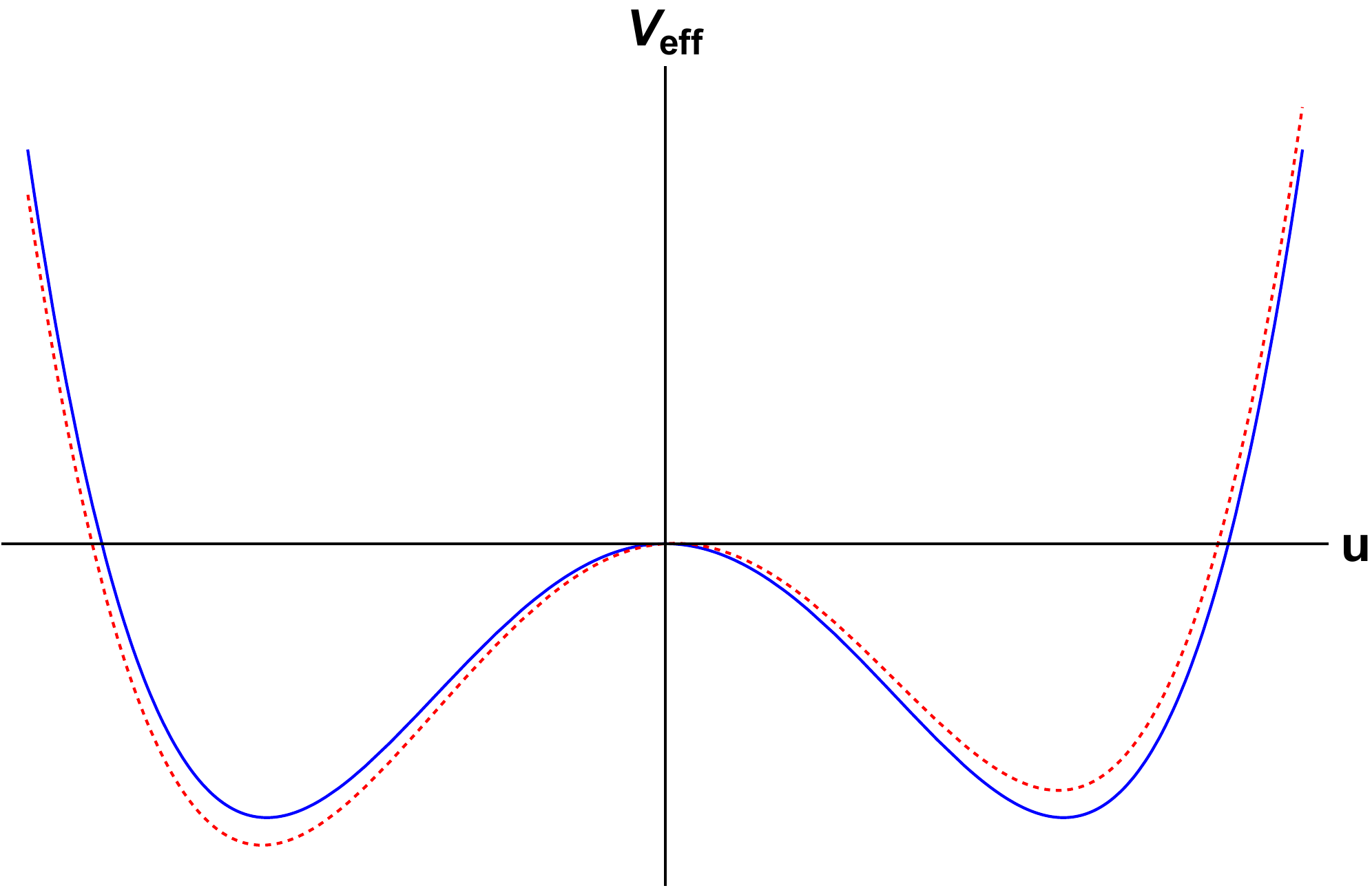}
		\caption{}\label{eff_pot_exp}
	\end{subfigure}
	\begin{subfigure}[b]{0.49\textwidth}
		\includegraphics[width=\textwidth]{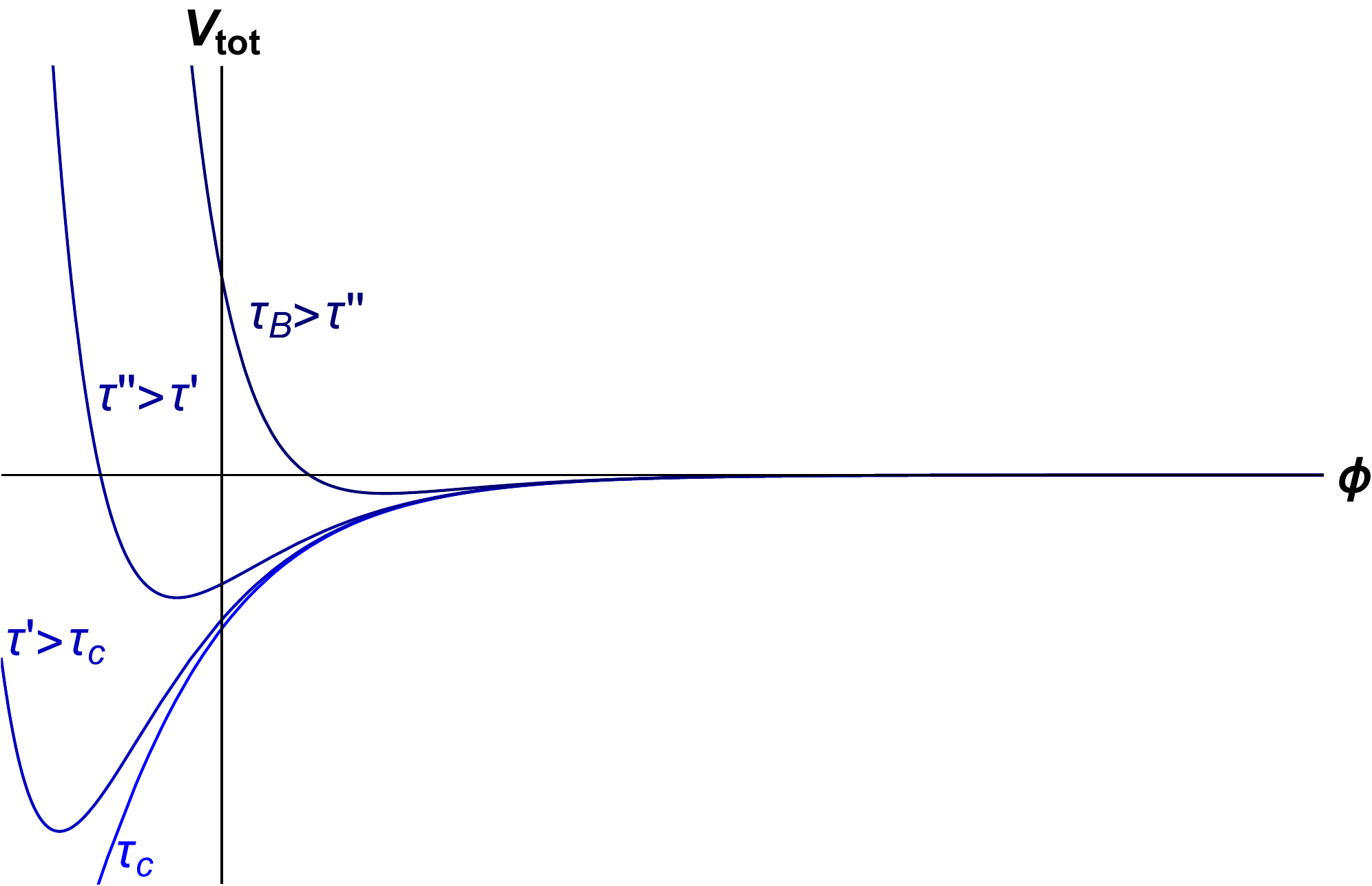}
		\caption{}\label{exp_to_morse}
	\end{subfigure}
	\caption{Left: NJL effective potential in the spontaneously broken phase. Solid blue: without explicit symmetry breaking, the effective potential is $\mathbb{Z}_2$ symmetric. Dashed red: with explicit symmetry breaking, the potential is tilted, and negative values of $u$ are favored.  \\
	Right: Total potential of the scalar field (\ref{total_pot}) at different times $\tau = \tau_c, < \tau', < \tau'',<\tau_B$. At $\tau <\tau_c$, $V_{tot}$ is just a negative exponent. After the chiral symmetry breaking, $V_{tot}$ transitions into the Morse potential ($n=2$), which has an exponentially increasing wall from which the scalar field bounces and continues to move towards increasing positive values.}
\end{figure}

As the background space-time is FRW after a period of slow contraction, the curvature is entirely determined by the time $\tau$, which has a negative value and is decreasing. At some moment $\tau_c$, the curvature will reach its critical value $R_c$, and the chiral symmetry will break. In our case, the condensate $u$ is a dynamical field and its evolution from $u=0$ to $u \neq 0$ is not spontaneous. We start the numerical evolution soon after the symmetry breaking, at $\tau_i  \gtrsim \tau_c ~(|\tau_c|  \gtrsim |\tau_i|)$. Before symmetry breaks, the solutions to the field equations are known analytically; see, for example, \cite{Weinberg:2008zzc} chapter 4.2. These solutions can be used to fix the initial values of $\varphi$ and $\varphi'$:
\bb\label{in_cond_phi}
\varphi (\tau_i) = \frac{1}{\kappa} \log \left[ \frac{\kappa^4}{2} ~\frac{\tau_i^2}{\kappa^2 - 6} \right];
\5
\varphi' (\tau_i) = \frac{2}{\kappa ~ \tau_i} ;
\ee
Before the chiral symmetry breaking, $u$ sits in the minimum of the effective potential, so $u(\tau )= u'(\tau) =0$ for $\tau \leq \tau_c$. To set initial conditions at $\tau = \tau_i$ the explicit breaking of the chiral symmetry needs to be accounted for. The explicit symmetry breaking (scalar-spinor mixing in (\ref{action_torsion_IO})) tilts the effective potential and forces condensate to move away from $u=0$, see Fig \ref{eff_pot_exp}. Ignoring the curvature corrections and accounting for the effect of explicit symmetry breaking within the effective potential $(\tau > \tau_c)$:
\bb
u '' (\tau) = -2 \xi \bar{F}_0 e^{-n \kappa \varphi} \left[ \log \left( \frac{e^{n \kappa \varphi}}{2 \xi \theta \bar{F}_0} \right) \right]^{-1} \equiv ddu (\tau) .
\ee
Since $\tau_i \approx \tau_c$, this equation can be approximately integrated and used to set the initial conditions for $u$:
\bb\label{in_cond_conden}
u(\tau_i) = \frac12 ddu(\tau_i) ~(\tau_i -\tau_c)^2 ;
\5\5\5
u'(\tau_i) = ddu(\tau_i) ~(\tau_i -\tau_c) .
\ee
To set up the initial condition for $h$, (\ref{Friedmann_num_form}) can be used, together with (\ref{in_cond_phi}) and (\ref{in_cond_conden}):
\bb
h(\tau_i) = - \sqrt{ \frac{4}{\kappa^4 ~\tau_i^2} - \frac{\bar{F}_0}{3} e^{-n \kappa \varphi (\tau_i) } u (\tau_i) - \frac{1}{12 \xi \theta } u ^2 (\tau_i) }
\ee

The scale standing in front of the negative exponent is chosen in the following way:
\bb
M_{pl}^2 V_0^2 = \left( 10^{-1} M_{pl} \right)^4 ;
\5\5\5 \rightarrow \5\5\5
V_0 = 10^{-2} M_{pl} .
\ee
In the numerical calculations $\kappa = 1000$. Since the value of $V_0$ is fixed, fixing the value of $\theta$ is equivalent to fixing the values of $\Lambda$ or $\xi_c$. In the calculations $\theta = 10^{-3}$ is used, which corresponds to $\Lambda =  M_{pl} / \sqrt{10}$ and $\xi_c = 2 \pi^2 \cdot 10 \approx 200$. The four-Fermi coupling, $\xi$, will be tuned to be just below its critical value: $\xi / \xi_c = 1-10^{-5}$. The IR cut-off scale $\epsilon = y \Lambda$ and explicit chiral symmetry breaking scale $F_0 \exp (- n \kappa \varphi (\tau_i))$ should be related and within an order of magnitude of each other:
\bb\label{expl_br_approx}
y \sim 2 \bar{F}_0 \xi \theta \exp (- n \kappa \varphi (\tau_i)) \ll 1 .
\ee
The last inequality ensures the validity of the approximation we took by ignoring the explicit symmetry breaking in Appendix \ref{app_loop_GR}. Explicitly:
\bb
y= 10^{-3} ;
\5\5\5
\frac{F_0}{\epsilon} \exp (- n \kappa \varphi (\tau_i)) = 10  ;
\5\5\5
\rightarrow
\bar{F}_0 \approx 92.38 .
\ee
The approximation (\ref{expl_br_approx}) breaks down and the value of $F_0 \exp (- n \kappa \varphi)$ becomes trans-Planckian when $\varphi$ becomes negative. This problem doesn't exist in \cite{Tukhashvili:2023itb}, where a rigid mass of the spinor was used.
Note that compared to \cite{Cook:2020oaj,Ijjas:2020dws,Ijjas:2021gkf,Ijjas:2021wml}, we had to add three additional parameters, two of them are associated with the NJL part of the action ($\theta; ~ \xi$) and one with the scalar-spinor mixing ($\bar{F}_0$). In this sense, the model is very minimal. We do not count $F(\tau_i)/\epsilon$ and $n$ as additional parameters, as their values do not affect the model's outcome. \\

\begin{figure}[t]
	\centering
	\begin{subfigure}[b]{0.49\textwidth}
		\includegraphics[width=\textwidth]{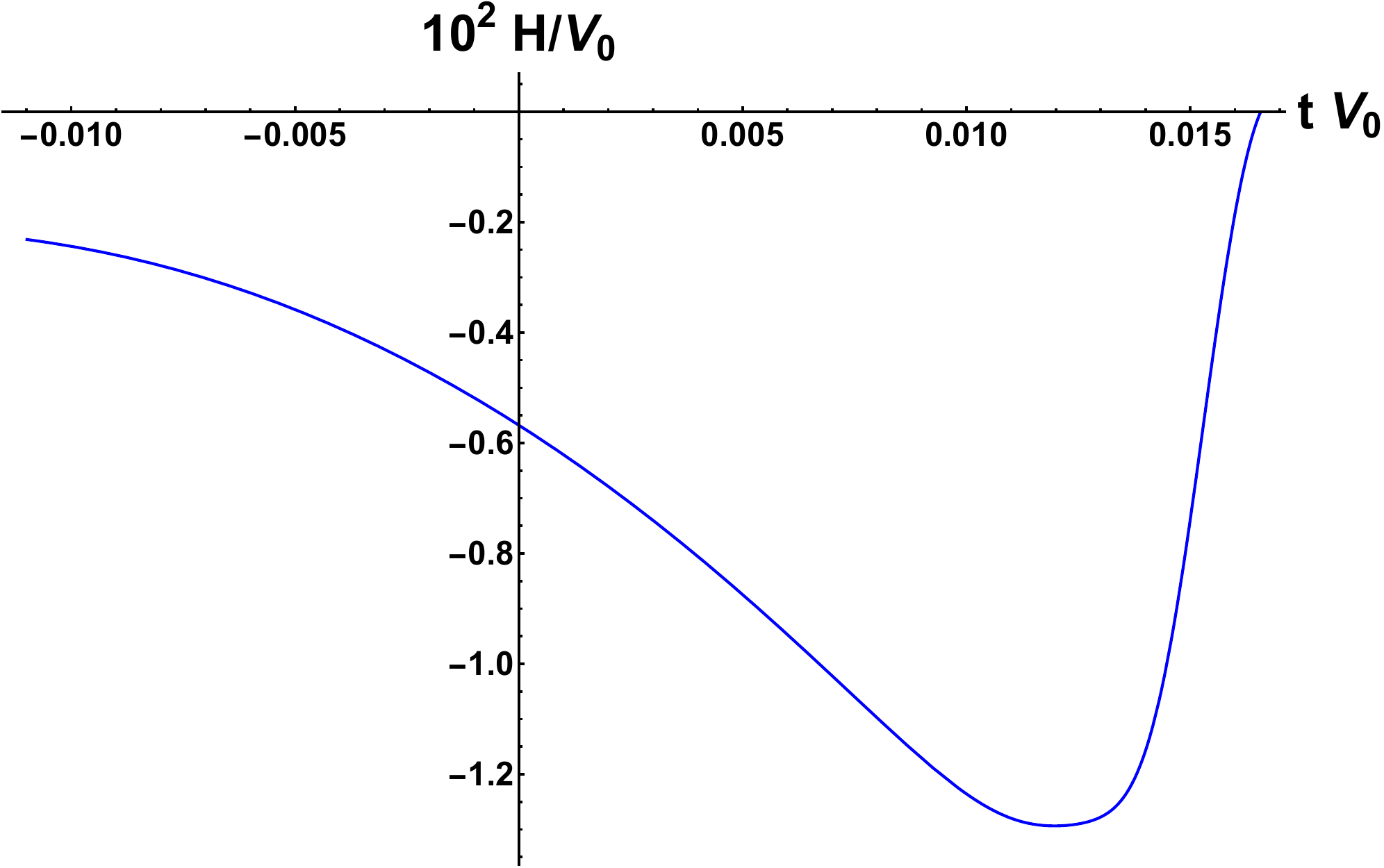}
		\caption{}\label{hubble.pdf}
	\end{subfigure}
	\begin{subfigure}[b]{0.49\textwidth}
		\includegraphics[width=\textwidth]{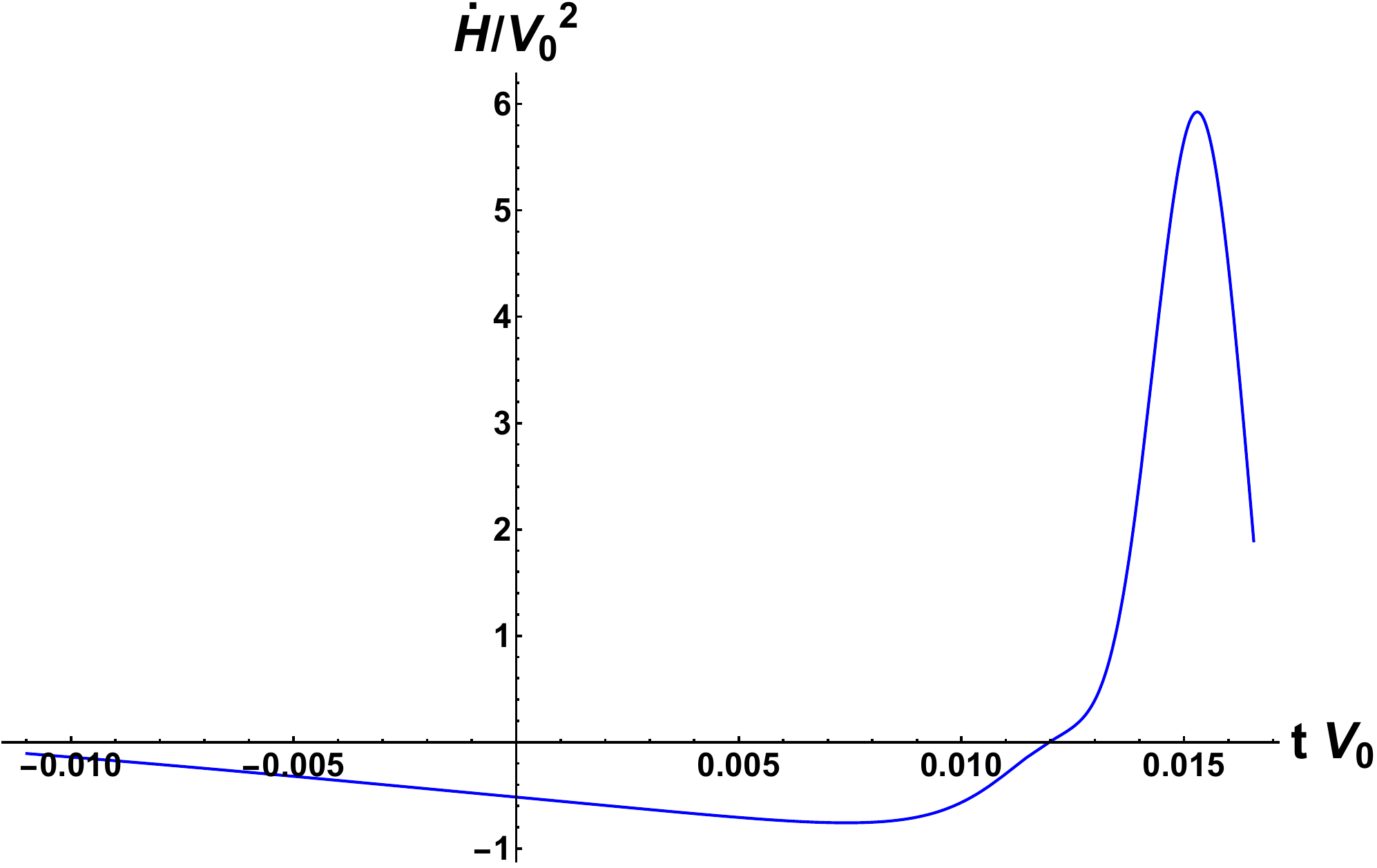}
		\caption{}\label{hubble_dot.pdf}
	\end{subfigure}
	\caption{Left: The Hubble parameter, $h = H/V_0$, as a function of $\tau=t V_0$. Right: Derivative of the Hubble parameter as a function of $\tau$. The violation of the NEC happens at around $\tau \approx 0.012 $. At this moment, the Hubble parameter starts to grow. At around $\tau \approx 0.015$, the derivative of the Hubble parameter reaches its maximum value and starts to decrease, but before $\dot{H}$ reaches zero, the bounce ($H=0$) happens.}
\end{figure}

\begin{figure}[t]
	\centering
	\begin{subfigure}[b]{0.49\textwidth}
		\includegraphics[width=\textwidth]{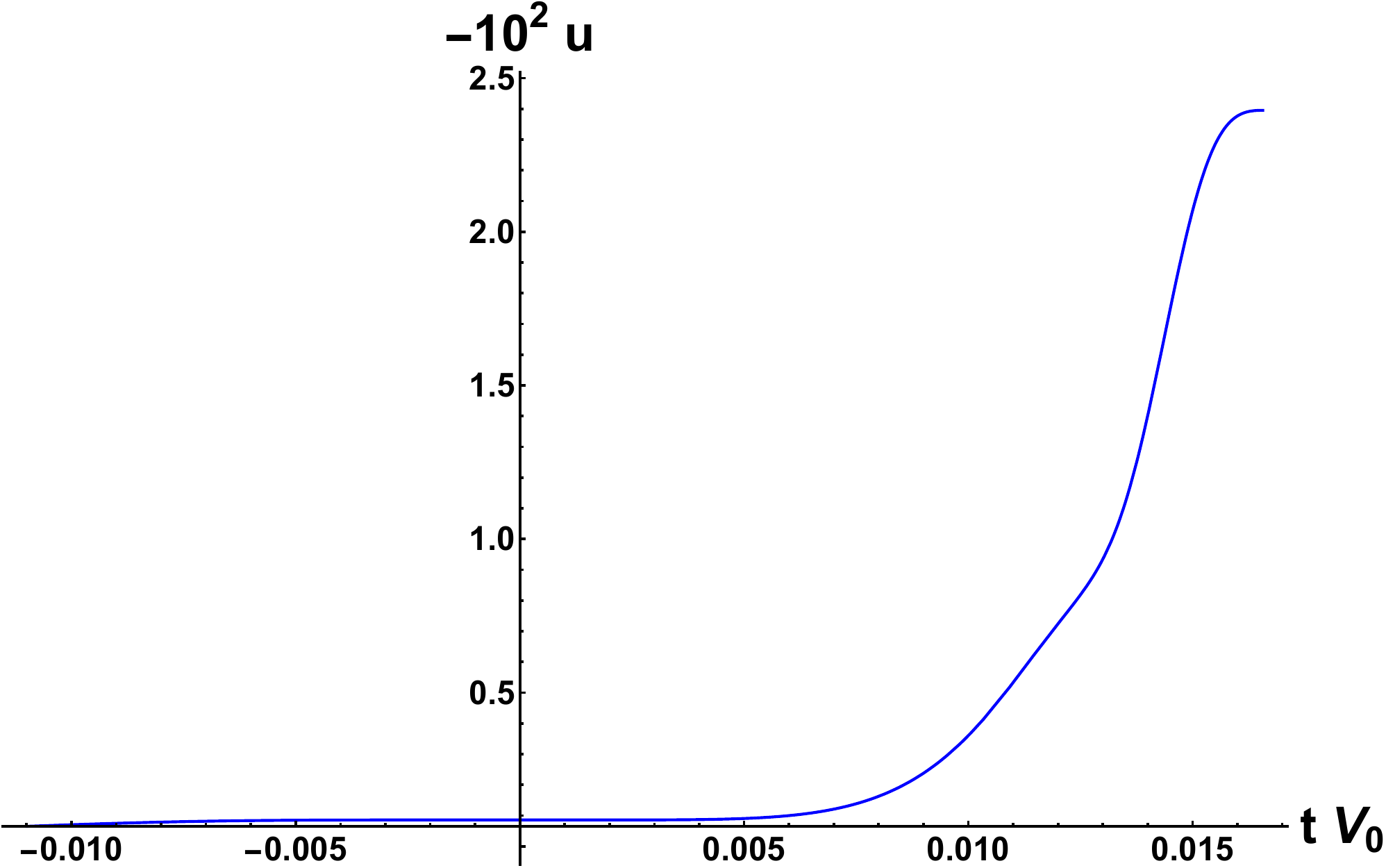}
		\caption{}\label{condensate.pdf}
	\end{subfigure}
	\begin{subfigure}[b]{0.49\textwidth}
		\includegraphics[width=\textwidth]{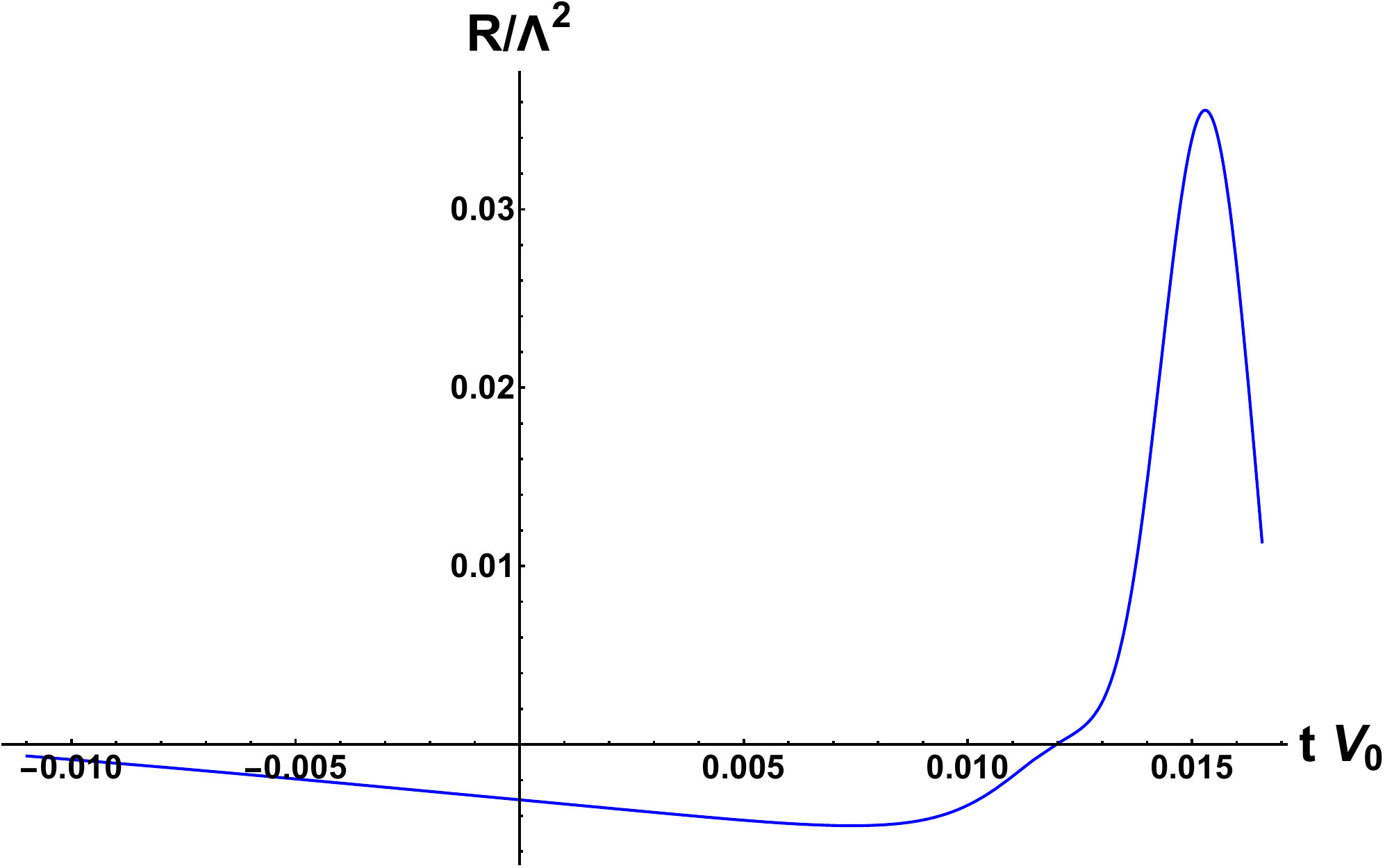}
		\caption{}\label{Ricci.pdf}
	\end{subfigure}
	\caption{Left: The condensate, $u$, as a function of $\tau$. After the chiral symmetry breaking, the condensate evolves with initial velocity defined by the explicit symmetry breaking (\ref{in_cond_conden}). Soon after this, the chiral symmetry gets restored, but due to the Hubble anti-friction, the absolute value of $u$ continues to increase. At around $\tau \approx 0.00125 $, the chiral symmetry breaks again and we observe a sharp rise of $u$. As $h \rightarrow 0$ chiral symmetry gets restored and the wall of the effective potential $u$ is climbing becomes steeper and steeper, this decelerates $u$; its velocity becomes zero at the moment of bounce, while $u' /h$ is finite. The latter guarantees finiteness of $h'$, as in any other case, it would blow up, see (\ref{h_dot_num_form}). \\ 
		Right: Ricci scalar curvature as a function of $\tau$. Its Maximum value is reached at around $\tau \approx 0.015$ and is $R \approx 0.04 \Lambda^2 = 0.004 M_{pl}^2$. This value is several orders below the Planck mass, so calling the bounce classical is justified, as the quantum gravity regime is never reached. It is also $\ll \Lambda^2$, so ignoring higher order curvature corrections in (\ref{eff_action_CC_kin}) is justified too. }
\end{figure}

\begin{figure}[b]
	\centering
	\begin{subfigure}[b]{0.49\textwidth}
		\includegraphics[width=\textwidth]{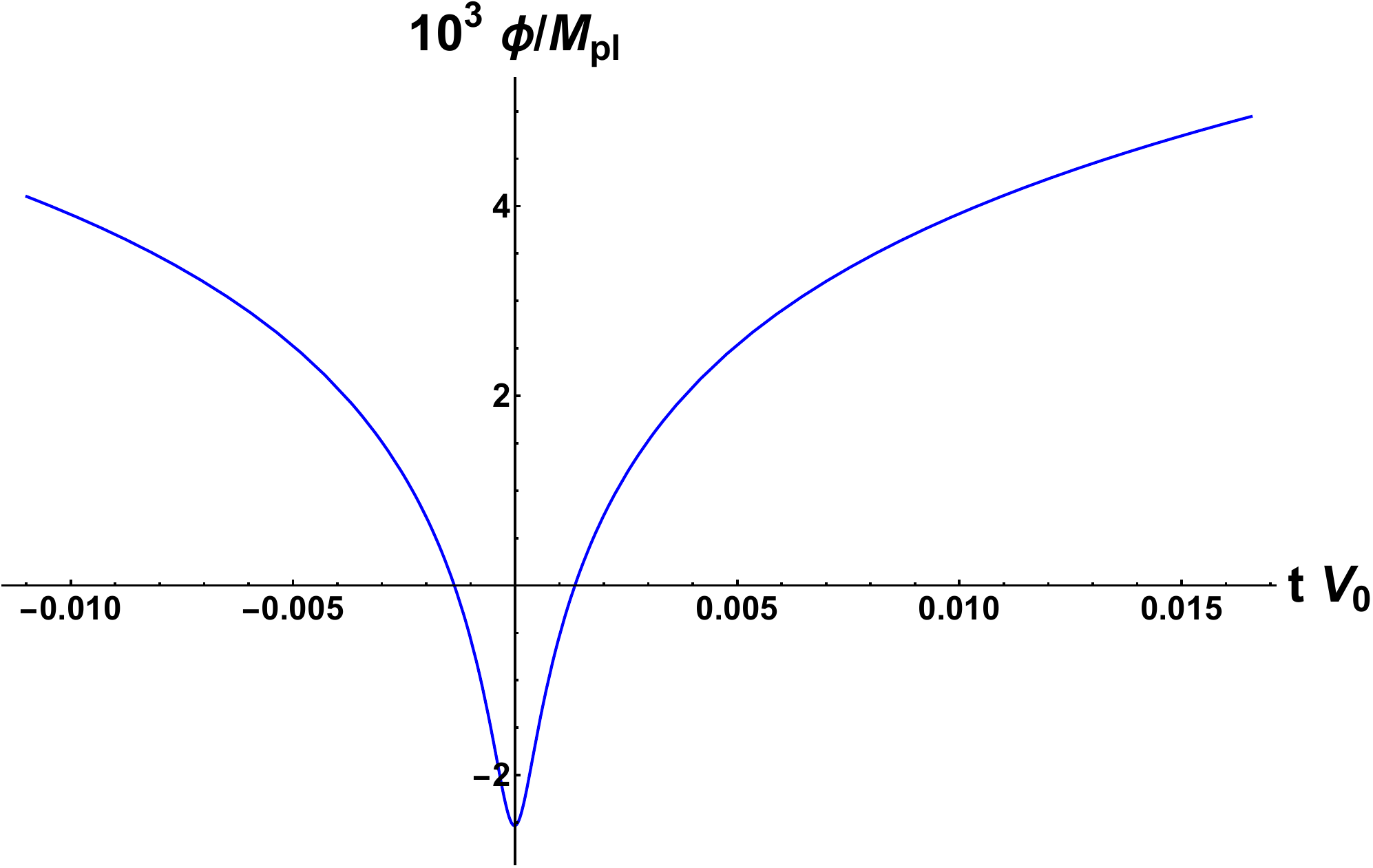}
		\caption{}\label{scalar.pdf}
	\end{subfigure}
		\begin{subfigure}[b]{0.49\textwidth}
		\includegraphics[width=\textwidth]{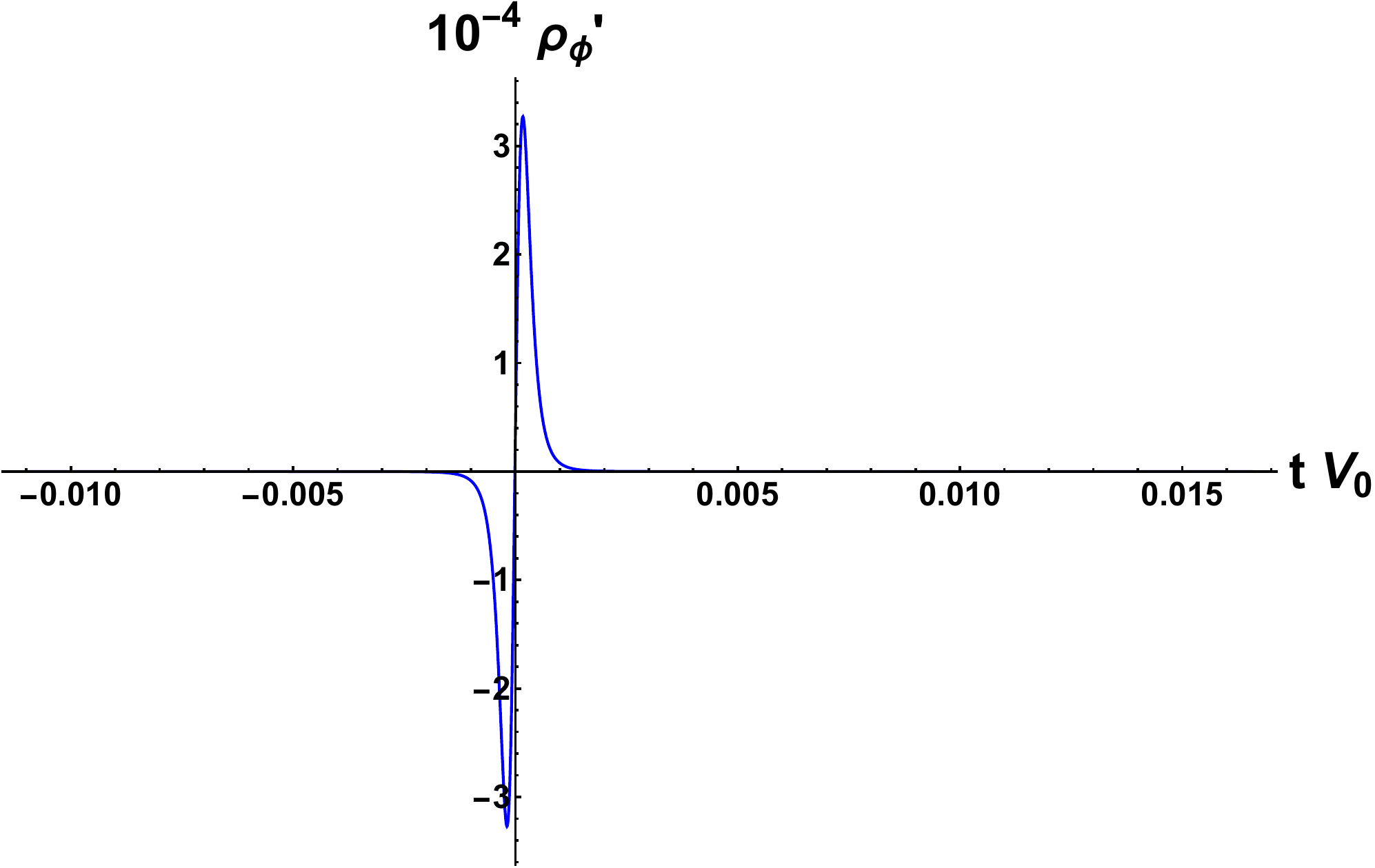}
		\caption{}\label{scalar_energy.pdf}
	\end{subfigure}
	\caption{Left: The scalar field $\varphi  = \phi /M_{pl}$ as a function of $\tau$. The dynamics is consistent with $\phi$ moving under the influence of the (\ref{total_pot}) potential. After the chiral symmetry breaking, the negative exponent effectively becomes a Morse potential, so the scalar field, $\phi$, which was moving from positive to negative values, finds an exponential wall and bounces from it. This happens at around $\tau \approx 0$. Then $\phi$ continues to move along the exponential plateau. After the bounce, it stops at some positive value due to the Hubble friction (not shown here). \\
		Right: Change of the scalar field energy density, $\rho_\phi$, as a function of $\tau$. The negative values of $\rho_\phi '$ follow from the fact that, as $\phi$ is climbing the exponential wall, it is decelerating. After it reaches the top of the wall and its velocity changes sign, $\phi$ starts to slide down and $\rho_\phi '$ becomes positive again. At the same time as $h\rightarrow 0$, the Hubble anti-friction is decreasing, so $\rho_\phi '$ approaches zero from positive values.}
\end{figure}

For the mentioned values of $\kappa$, $\xi$, $y$, and $\theta$, the chiral symmetry breaking happens at approximately $\tau_c = -0.0355$. The choice of $\tau_i$ does affect the outcome of the model. Even though it does not affect the violation of NEC, it does affect the bounce. We solved the system (\ref{scalar_num_form})-(\ref{condensate_num_form}) for hundred $\tau_i$ evenly distributed between the interval $[-0.0119,-0.0109]$ (chosen randomly). We observed the NEC violation for all values of $\tau_i$, but the bounce took place only in five cases. 
If $\tau_i$ is chosen randomly at different patches of space, those patches that bounce will dominate the volume as they are the ones expanding. In contrast, those patches that do not bounce will continue to shrink and eventually crunch to a point.
One of the choices for which the bounce does happen is $\tau_i = -0.01099$. The numerical solutions for the dimensionless Hubble parameter $h$, its derivative $h'$, the condensate $u$, the Ricci scalar curvature $R$, the scalar field $\varphi$ and the derivative of the scalar field energy density $\partial_\tau \left( \frac12 (\varphi' )^2- \exp(- \kappa \varphi) \right)$ are shown on Figs. \ref{hubble.pdf}, \ref{hubble_dot.pdf}, \ref{condensate.pdf}, \ref{Ricci.pdf}, \ref{scalar.pdf} and \ref{scalar_energy.pdf} respectively. 

At the moment of bounce, the velocity of the condensate must be zero, $\dot{u} = 0$. This is indeed the case as in Fig. \ref{condensate.pdf}. The reason is that derivatives of the Hubble parameter contain terms proportional to $\dot{u}/H$, and their signs are such that their large values favor chiral symmetry restoration. As an example for the dimensionless $\dot{H}$ we have:
\bb\label{h_dot_num_form}
h' = - \frac12 \left( \varphi' \right)^2 - \frac{u'}{6 h} \left( \frac{1}{2 \xi \theta} u + \bar{F}_0 e^{-n \kappa \varphi} \right) .
\ee
As $H \rightarrow 0$, derivatives of the Hubble parameter grow, and the wall of the effective potential $u$ is climbing becomes steeper. At the moment of bounce, $\tau_B$, $\dot{u}=0$, but $\dot{u}/H$ is finite, so the process is self-consistent, and singularity is never reached.

The scalar field effectively moves through the following potential:
\bb\label{total_pot}
V_{tot} = M_{pl}^2 V_0^2 \Big( - e^{- \kappa \varphi} - \bar{F}_0 u ~ e^{- n \kappa \varphi} \Big) .
\ee
When the chiral symmetry is unbroken, $u=0$, and $V_{tot}$ coincides with the negative exponential potential. Once the chiral symmetry breaks, $V_{tot}$ effectively becomes a Morse potential ($n=2$). Fig. \ref{exp_to_morse} shows $V_{tot}$ at different moments of the evolution. After the symmetry breaks, the total potential of the scalar becomes bounded from below and exponentially increasing for negative values of $\phi$. $\phi$, which before the chiral symmetry breaking was moving on a negative exponential potential from positive to negative values, after the chiral symmetry breaking encounters an exponential wall, bounces from it and continues to move on a plateau see Fig. \ref{scalar.pdf}. After the bounce, the field $\phi$ will decelerate due to the Hubble friction and will freeze at some positive value of $\phi$. 
Before the chiral symmetry breaking, the energy density of the scalar field, $\rho_\varphi$, is increasing due to the Hubble anti-friction. From Eq. (\ref{scalar_num_form}) it follows:
\bb\label{scalar_energy_change}
\rho_\varphi ' = \partial_\tau \left( \frac12 (\varphi' )^2- \exp(- \kappa \varphi) \right) =
- 3 h \left( \varphi' \right)^2 -  n \kappa \bar{F}_0 \varphi' e^{- n \kappa \varphi} u .
\ee
The first term on the RHS of (\ref{scalar_energy_change}) is positive during the contraction and becomes zero at the moment of bounce, while the sign of the second term depends on the sign of $\varphi'$. After chiral symmetry breaking, while the scalar is climbing the exponential wall, its velocity is negative and decreasing. Since $u$ is also negative, the second term contributes negatively. As a result $\rho_\varphi$ decreases, see Fig \ref{scalar_energy.pdf}. One can argue that the scalar-condensate mixing causes effect of Hubble friction (deceleration of $\varphi$), which might be relevant in the production of the adiabatic perturbations \cite{Ijjas:2021ewd,Ijjas:2020cyh}. Keep in mind that, in our case, the deceleration happens not during the slow contraction but after that and just before the bounce, so the deceleration of the scalar field by itself might not be sufficient to achieve that goal.

During the bounce, the Hubble parameter $h = 0$ and the total energy density $\rho = 0$, the most reliable candidate to control the validity of the effective theory is the Ricci scalar curvature. From Fig. \ref{Ricci.pdf}, we see that the maximum value of Ricci curvature during the evolution is $R_{max} \approx 0.04 \Lambda^2 = 0.004 M_{pl}^2$, which is small enough for both the curvature expansion in (\ref{eff_action_CC_kin}) to be valid and gravity to stay in the classical regime. \\

%%%%%%%%%%%%%%%%%%%%%%%%%%%%%%%%%%%%%%%%%%%%%%%%%%%%%%%%%%%%%%%%%%%%%%%%%%
%%%%%%%%%%%%%%%%%%%%%%%%%%%%%%%%%%%%%%%%%%%%%%%%%%%%%%%%%%%%%%%%%%%%%%%%%%
%%%%%%%%%%%%%%%%%%%%%%%%%%%%%%%%%%%%%%%%%%%%%%%%%%%%%%%%%%%%%%%%%%%%%%%%%%

\section{\Large Discussion and Unanswered Questions}

%%%%%%%%%%%%%%%%%%%%%%%%%%%%%%%%%%%%%%%%%%%%%%%%%%%%%%%%%%%%%%%%%%%%%%%%%%
%%%%%%%%%%%%%%%%%%%%%%%%%%%%%%%%%%%%%%%%%%%%%%%%%%%%%%%%%%%%%%%%%%%%%%%%%%
%%%%%%%%%%%%%%%%%%%%%%%%%%%%%%%%%%%%%%%%%%%%%%%%%%%%%%%%%%%%%%%%%%%%%%%%%%

In this paper, we have extended our proposal \cite{Tukhashvili:2023itb} that a chiral condensate might induce a cosmological bounce. We explicitly showed that, under certain conditions, the Nambu-Jona-Lasinio condensation can be triggered with a gravitational field when the absolute value of the Ricci curvature exceeds a certain value. The NJL condensation is a vacuum effect and preserves the symmetries of space-time. This is a significant improvement compared to earlier works in the literature that attempted to use a dense gas of spin-aligned fermions to induce a bounce, which can destroy the FRW symmetries that must be preserved. Using the numerical solutions, it was demonstrated that this condensate can act as the source of NEC violation and ultimately lead to a cosmological bounce. We also addressed several points not discussed in \cite{Tukhashvili:2023itb}; these include the derivation of the initial conditions for the condensate and derivation of the NJL effective action in the presence of a gravitational field. Compared to \cite{Tukhashvili:2023itb}, a different potential for the scalar field was used. The fact that the mechanism works in both cases hints at its robust nature. We have yet to address what might happen after the bounce or the perturbative stability of the solutions. This will be the subject of future works. 

The four-Fermi interactions that cause condensation create an attractive force between the spinors. Gravity always creates attractive force, so including gravitational effects should always enhance the condensation process. According to the analysis and arguments of section \ref{Q_Average}, this is not the case, as a positive Ricci curvature can restore the chiral symmetry even when NJL coupling constant exceeds its (flat space) critical value. One of the mysteries that will be explored in future work is explaining how gravity can act to prevent chiral symmetry breaking. 

%%%%%%%%%%%%%%%%%%%%%%%%%%%%%%%%%%%%%%%%%%%%%%%%%%%%%%%%%%%%%%%%%%%%%%%%%%
%%%%%%%%%%%%%%%%%%%%%%%%%%%%%%%%%%%%%%%%%%%%%%%%%%%%%%%%%%%%%%%%%%%%%%%%%%
%%%%%%%%%%%%%%%%%%%%%%%%%%%%%%%%%%%%%%%%%%%%%%%%%%%%%%%%%%%%%%%%%%%%%%%%%%

\section*{\Large Acknowledgements}

I would like to express my gratitude to Paul Steinhardt for his invaluable support and guidance throughout this project.
I would like to thank Greg Gabadadze, Anna Ijjas, Otari Sakhelashvili, Richard Woodard, David Shlivko, Stephon Alexander, Andrew Sullivan, Alejandro C\'ardenas-Avenda\~no for discussions and comments; Thibault Damour, 
Alexander Vikman and Ilya Shapiro for correspondence.
This work is supported by the Simons Foundation grant number 654561.

%%%%%%%%%%%%%%%%%%%%%%%%%%%%%%%%%%%%%%%%%%%%%%%%%%%%%%%%%%%%%%%%%%%%%%%%%%
%%%%%%%%%%%%%%%%%%%%%%%%%%%%%%%%%%%%%%%%%%%%%%%%%%%%%%%%%%%%%%%%%%%%%%%%%%
%%%%%%%%%%%%%%%%%%%%%%%%%%%%%%%%%%%%%%%%%%%%%%%%%%%%%%%%%%%%%%%%%%%%%%%%%%

\appendix

%%%%%%%%%%%%%%%%%%%%%%%%%%%%%%%%%%%%%%%%%%%%%%%%%%%%%%%%%%%%%%%%%%%%%%%%%%
%%%%%%%%%%%%%%%%%%%%%%%%%%%%%%%%%%%%%%%%%%%%%%%%%%%%%%%%%%%%%%%%%%%%%%%%%%
%%%%%%%%%%%%%%%%%%%%%%%%%%%%%%%%%%%%%%%%%%%%%%%%%%%%%%%%%%%%%%%%%%%%%%%%%%

\section{\Large Integrating out the Torsion}\label{app_int_tor}

First of all, let's split the action (\ref{orig_action}) into torsion dependent, $\mathcal{S}_T$, and torsion independent, $\mathcal{S}_0$, parts:
\bb\label{app_action}
\mathcal{S} = \mathcal{S}_0 + \mathcal{S}_T ;
\ee
\begingroup\makeatletter\def\f@size{11}\check@mathfonts
\begin{align}\label{tor_action}
	\nonumber   \mathcal{S}_0 = &  - \frac{M_{pl}^2}{4} \epsilon_{A B M N}
	e^A \wedge e^B \wedge R^{M N} - \frac{i}{2 \cdot 3!} \epsilon_{A B M N} e^A \wedge e^B \wedge e^M \wedge  \Big( \overline{\Psi} \gamma^N D \Psi - D \overline{\Psi} \gamma^N \Psi \Big)   \\
	\nonumber {} & + \frac{1}{4!}  \epsilon_{A B M N} e^A \wedge e^B \wedge e^M \wedge e^N
	\left[ \frac12 (\partial \phi)^2 + V (\phi) + F ( \phi ) \overline{\Psi} \Psi \right] ; \\
	\nonumber  \mathcal{S}_T = & - \frac{M_{pl}^2}{4} \epsilon_{A B M N}
	e^A \wedge e^B \wedge C^{M K} \wedge {C_K}^N  - \frac{1}{4!} \epsilon_{A B M N} {\epsilon_{I J }}^{N K} e^A \wedge e^B \wedge e^M \wedge C^{I J} A_K  \\
	{} & + \sqrt{\frac{\xi}{6}} \epsilon_{I J K L} {C^I}_M \wedge e^M \wedge e^J \wedge e^K ~V^L + \left( \frac{1}{4} - \sqrt{\frac{\xi}{6}} \right) {C^I}_K \wedge e^K \wedge e_I \wedge e_J ~A^J .
\end{align}
\endgroup
Expressions without the tilde are defined wrt $\omega^{AB}$ and satisfy the following Cartan's structure equations:
\bb
D e^A = d e^A + \left. \omega^A \right.{}_B \wedge e^B = 0 ;
\ee
\bb
0 = D D e^A = \left. R^A \right.{}_B \wedge e^B .
\ee
Relation between the different spin connections and curvature two forms:
\bb
\tilde{\omega}^{AB} = \omega^{AB} + C^{AB} ;
\5\5\5
\tilde{R}^{AB} = R^{AB} + D C^{AB} + C^{A K} \wedge {C_K}^B .
\ee
Since $C^{AB}$ is a difference between the connections, it is a tensor. Variation of the $\mathcal{S}_T$ wrt to $C$, yields an algebraic equation:
\begingroup\makeatletter\def\f@size{11}\check@mathfonts
\bb\label{gant}
{C^X}_{MN} - {C^X}_{NM} + \delta^X_M {C^K}_{NK} - \delta^X_N {C^K}_{MK} + \frac{1}{M_{pl}^2} \sqrt{\frac{2 \xi}{3}} \Big( \delta^X_M V_N - \delta^X_N V_M - {\epsilon^X}_{MNK} A^K \Big) = 0
\ee
\endgroup
Tracing this equation gives:
\bb\label{gant_tr}
{C^K}_{MK} = - \frac{1}{ M_{pl}^2} \sqrt{\frac{3 \xi}{2}} V_M .
\ee
The following ansatz is chosen for $C$:
\bb\label{ansatz}
{C^X}_{MN} = {\epsilon^X}_{MNK} P^K + \delta^X_N Q_M - \eta_{MN} Q^X .
\ee
Note that this is not the most general ansatz for $C$. There should also be a 3-index term \cite{Bagrov:1994re}. From (\ref{gant}), we see that nothing can source the 3-index term, so we set it to zero from the beginning. Using (\ref{ansatz}) into (\ref{gant_tr}) gives:
\bb\label{tor_sol_Q}
Q_M =  - \frac{1}{M_{pl}^2} \sqrt{\frac{ \xi}{6}} V_M .
\ee
After plugging this into (\ref{gant}):
\bb\label{tor_sol_P}
P_M = \frac{1}{M_{pl}^2} \sqrt{\frac{ \xi}{6}} A_M  .
\ee
Using (\ref{tor_sol_Q}) and (\ref{tor_sol_P}) into (\ref{ansatz}) and plugging the latter into (\ref{tor_action}), the torsion-dependent part of the action can be expressed in terms of spinor currents:
\bb
\mathcal{S}_T = - \frac{1}{4!} \epsilon_{A B I J} e^A \wedge e^B \wedge e^I \wedge e^J \left[
\frac{\xi}{2 M_{pl}^2} \left( V_M V^M - A_M A^M \right) \right] .
\ee
The latter can be further simplified with the help of the Fierz identity (\ref{Fierz_id}). The final expression for the torsion-dependent part:
\bb
\mathcal{S}_T = - \frac{\xi}{4! M_{pl}^2} \epsilon_{A B I J} e^A \wedge e^B \wedge e^I \wedge e^J
\left[ \big( \overline{\Psi} \Psi \big)^2 + \big( i \overline{\Psi} \gamma_5 \Psi \big)^2 \right]
\ee
does not involve any spinor current vectors, just the scalar and the pseudo-scalar bi-linears.

%%%%%%%%%%%%%%%%%%%%%%%%%%%%%%%%%%%%%%%%%%%%%%%%%%%%%%%%%%%%%%%%%%%%%%%%%%
%%%%%%%%%%%%%%%%%%%%%%%%%%%%%%%%%%%%%%%%%%%%%%%%%%%%%%%%%%%%%%%%%%%%%%%%%%
%%%%%%%%%%%%%%%%%%%%%%%%%%%%%%%%%%%%%%%%%%%%%%%%%%%%%%%%%%%%%%%%%%%%%%%%%%

\section{\Large Fermion Loop in External Gravitational Field}\label{app_loop_GR}

%%%%%%%%%%%%%%%%%%%%%%%%%%%%%%%%%%%%%%%%%%%%%%%%%%%%%%%%%%%%%%%%%%%%%%%%%%
%%%%%%%%%%%%%%%%%%%%%%%%%%%%%%%%%%%%%%%%%%%%%%%%%%%%%%%%%%%%%%%%%%%%%%%%%%
%%%%%%%%%%%%%%%%%%%%%%%%%%%%%%%%%%%%%%%%%%%%%%%%%%%%%%%%%%%%%%%%%%%%%%%%%%

This appendix aims to calculate the spinor loop corrections due to the Gravitational field's presence up to the second order in the Riemann curvature tensor. The Riemann Normal Coordinates are used to achieve this goal, but in the third part of this appendix, the same calculation is done using the more rigorous Heat Kernel method, and the two results are compared.

%%%%%%%%%%%%%%%%%%%%%%%%%%%%%%%%%%%%%%%%%%%%%%%%%%%%%%%%%%%%%%%%%%%%%%%%%%

\subsection{\large Riemann Normal Coordinates}

%%%%%%%%%%%%%%%%%%%%%%%%%%%%%%%%%%%%%%%%%%%%%%%%%%%%%%%%%%%%%%%%%%%%%%%%%%

Given a curved space-time, one can always find inertial frames at a given event $\mathcal{P}_0$. Free particles move along straight lines within these frames, meaning $\Gamma^\rho_{\mu \nu} =0$. This all is true only locally. The geodesics coming out of $\mathcal{P}_0$ can be used to define the neighboring points. This is the main idea in defining the Riemann normal coordinates. The Riemann normal coordinates $x^\mu$ can be chosen in a way that $x^\mu (\mathcal{P}_0) = 0$ and $g_{\mu \nu} (\mathcal{P}_0) = \eta_{\mu \nu}$. In addition if  $\mathcal{P}$ is a point in the neighborhood of $\mathcal{P}_0$, then $x^\mu (\mathcal{P}) = u p^\mu$ where $ p^\mu = d x^\mu /du (\mathcal{P}_0) $ is the tangent vector at $\mathcal{P}_0$ to the geodesic $\mathcal{P}_0 \mathcal{P}$ and $u$ is the proper time (assuming $\mathcal{P}_0 \mathcal{P}$ is time-like) \cite{Pirani:1956wr}. Using these properties, one can expand the metric at $\mathcal{P}$ in ``powers" of the Riemann curvature tensor \cite{Petrov:1969,Misner:1973prb,Brewin:1996yk}:
\begin{align}\label{metric_RNC}
	g_{\mu \nu} (\mathcal{P}) = & ~\eta_{\mu \nu} - \frac{1}{3} R_{\mu \alpha \nu \beta} x^\alpha x^\beta
	- \frac{1}{6} \nabla_\gamma R_{\mu \alpha \nu \beta} x^\alpha x^\beta x^\gamma  \\
	\nonumber {} & + \left( - \frac{1}{20} \nabla_\gamma  \nabla_\delta R_{\mu \alpha \nu \beta}
	+ \frac{2}{45} R_{\alpha \mu \beta \lambda} {R^\lambda}_{\gamma \nu \delta} \right)
	x^\alpha x^\beta x^\gamma x^\delta + \mathcal{O} \Big( \left( \partial g \right)^5 \Big) .
\end{align}
From now on, I will suppress $\mathcal{P}$ from the left-hand side. Having the expression for the metric (\ref{metric_RNC}), one can evaluate the Christoffel gamma, the spin connection, the tetrad, the determinant of the metric, etc., as a series in the Riemann curvature tensor. Obtaining these expressions is a tedious but straightforward task and will not be shown here due to their length. \\

Calculation of the effective action for spinors involves evaluation of the functional trace $(\text{Tr})$ from the spinor two-point function $S$:
\bb
\text{Tr} S (x,z,s) = \int d^4 x   \sqrt{-g}~ \lim_{x \rightarrow z} ~\text{tr} S (x,z,s) .
\ee
Here $\text{tr}$ stands for the trace over the Dirac, flavor, color, etc. indices. Since $S (x,z,s)$ depends only on the difference $x-z$, and since this procedure involves taking the limit $x \rightarrow z $, using Riemann normal coordinates to evaluate the effective action is perfectly justified. \\

Let's focus on the spinor part of the action and ignore the terms that explicitly break the chiral symmetry by making the usual assumptions. The relevant part of the action:
\bb
\mathcal{S}_{NJL} = \int d^4 x \sqrt{-g} \left[ i \overline{\Psi} \gamma^\mu \nabla_\mu \Psi + \frac{\xi}{M_{pl}^2}
\Big( \left( \overline{\Psi} \Psi \right)^2 + \left( i \overline{\Psi} \gamma_5 \Psi \right)^2 \Big)  \right] ,
\ee
with $\gamma^\mu = e^\mu_A \gamma^A$. By introducing the auxiliary fields:
\bb\label{aux_field_def}
\sigma =  - \frac{2 \xi}{M_{pl}^2} \overline{\Psi} \Psi ;
\5\5\5
\pi = - \frac{2 \xi}{M_{pl}^2} \left( i \overline{\Psi} \gamma_5 \Psi \right) ,
\ee
one can rewrite this action as:
\bb\label{actio_aux}
\mathcal{S}_{NJL} = \int d^4 x \sqrt{-g} \left[ \overline{\Psi} \Big( i \gamma^\mu \nabla_\mu -\sigma - i \pi \gamma_5 \Big) \Psi - \frac{M_{pl}^2}{4 \xi} \Big( \sigma^2 + \pi^2 \Big)  \right] .
\ee
The chiral symmetry allows us to set $\pi = 0$. After integrating out the spinor fields, the effective action is obtained:
\bb
\int \mathcal{D} \sigma \mathcal{D} \pi \mathcal{D} \overline{\Psi} \mathcal{D} \Psi
\exp \Big( i~ \mathcal{S}_{NJL} \Big) =
\int \mathcal{D} \sigma \mathcal{D} \pi \exp \Big( i~ \mathcal{S}_{eff} \Big) ;
\ee
\bb
\mathcal{S}_{eff} = \int d^4 x \sqrt{-g} \left[ i  \int_0^\sigma ds ~\text{tr} S (x,x, s) - \frac{M_{pl}^2}{4 \xi} \sigma^2  \right] .
\ee
A few words need to be said about the scalar $\sigma$. When the space-time is uniform and static, so must be $\sigma$, but since one is dealing with a dynamical space-time $\partial \sigma \neq 0$. For now, let's (wrongly) assume that $\sigma$ is uniform and static. The corrections involving non-zero $\partial \sigma$ will be accounted for later.

The spinor Greens function in a curved space-time satisfies the following equation:
\bb
\big( i  \gamma^\mu \nabla_\mu - s \big) S (x,z, s) = \frac{1}{\sqrt{g}} \delta^{(4)} (x-z) .
\ee
Let's define an auxiliary function:
\bb
\big( i  \gamma^\mu \nabla_\mu + s \big) F (x,z, s) = S (x,z, s) ,
\ee
satisfying:
\bb\label{squared_green}
\sqrt{-g} \left( g^{\mu \nu} \nabla_\mu \nabla_\nu - \frac{1}{4} R - s^2 \right) F (x,z, s) = \delta^{(4)} (x-z) .
\ee
At this point, one can use the Riemann normal coordinates to expand the LHS in the series of Riemann curvatures and solve this equation iteratively in the Fourier space. After a lengthy calculation we arrive \cite{Bunch:1979uk,Parker:1983pe}:
\begin{align*}
	F (p) = & -\frac{1}{p^2 + s^2 } - \frac{1}{12} \frac{R}{(p^2 + s^2)^2} + \frac{2}{3} \frac{R^{\mu \nu} p_\mu p_\nu}{(p^2 + s^2)^3} \\
	{} & + \frac{i}{2} \frac{\nabla_\alpha R ~p^\alpha}{(p^2 + s^2)^3}
	- \frac{i}{12} \frac{\nabla_\alpha \mathcal{R}^{\alpha \beta} ~p_\beta }{(p^2 + s^2)^3}
	- 2 i \frac{\nabla_\gamma R_{\alpha \beta} ~p^\alpha p^\beta p^\gamma}{(p^2 + s^2)^4} \\
	{} & - \frac{1}{(p^2 + s^2)^3} \left( \frac{3}{20} \nabla^2 R + \frac{1}{144} R^2
	+ \frac{2}{45} R_{\mu \nu} R^{\mu \nu}
	+ \frac{1}{15} R_{\mu \nu \rho \sigma} R^{\mu \nu \rho \sigma}
	+ \frac{1}{128} \mathcal{R}_{\mu \nu} \mathcal{R}^{\mu \nu} \right) \\
	{} & + \frac{1}{(p^2 + s^2)^4} \left(  \frac{4}{5} \nabla^2 R_{\alpha \beta}
	+ \frac{7}{5} \nabla_\alpha \nabla_\beta R
	+ \frac{1}{6} R R_{\alpha \beta} + \frac{8}{5} R_{\alpha \rho} R^\rho_\beta
	- \frac{4}{5} R^{\mu \nu} R_{\mu  \alpha \nu \beta} \right. \\
	\nonumber {} & ~~~~~~~~~~~~~~~~~~~ \left. + \frac{8}{15} {R^{\mu \nu \rho}}_\alpha R_{\mu \nu \rho \beta}
	- \frac{1}{4} \nabla_\alpha \nabla_\rho {\mathcal{R}^\rho}_\beta
	+ \frac{1}{32} \mathcal{R}_{\mu \alpha} {\mathcal{R}^\mu}_\beta \right) p^\alpha p^\beta \\
	\nonumber {} & - \frac{1}{(p^2 + s^2)^5} \left( \frac{24}{5} \nabla_\gamma \nabla_\delta R_{\alpha \beta}
	+ \frac{16}{15} R_{\mu \alpha \nu \beta} {{{R^\mu}_\gamma}^\nu}_\delta
	+ \frac{4}{3} R_{\alpha \beta} R_{\gamma \delta} \right) p^\alpha p^\beta p^\gamma p^\delta ,
\end{align*}
where:
\bb
\mathcal{R}_{\mu \nu} \equiv R_{A B \mu \nu} \left[ \gamma^A , \gamma^B \right] .
\ee
Using the auxiliary function $F(p)$, the effective action can be written as:
\bb
\mathcal{S}_{eff} = \int d^4 x \sqrt{-g} \left[ i \int_0^\sigma ds ~s \int \frac{d^4 p}{(2 \pi)^4} ~ \text{tr} F(p) - \frac{M_{pl}^2}{4 \xi} \sigma^2  \right] .
\ee
From this point, one can proceed as follows. First, symmetrize the integrals by making the replacements (odd powers of $p$ do not contribute to the integration):
\bb
p^\alpha p^\beta \rightarrow \frac{p^2}{4} \eta^{\alpha \beta} ;
\5\5\5
p^\alpha p^\beta p^\gamma p^\delta \rightarrow \frac{p^4}{4 (4+2)}
\Big( \eta^{\alpha \beta} \eta^{\gamma \delta} + \eta^{\alpha \gamma} \eta^{\beta \delta}
+ \eta^{\alpha \delta} \eta^{\beta \gamma} \Big) .
\ee
After which the Wick rotation, $p_0 \rightarrow i p_{0E}$, is done, and the momentum integrals are cut at the UV scale $p_E^2 = \Lambda^2$ and IR scale $p_E^2 = \epsilon^2$. Then the final integration wrt $s$ is performed. The flat space invariant cut-off is used because the curvature is treated as a perturbation over the flat space. The necessity of the IR scale is discussed in section \ref{Q_Average}. Finally, the effective action involving corrections up to the second order in curvature is obtained:
\begin{align}\label{eff_action_CC}
	\nonumber {} & S_{eff}^{CC} = \frac{ \Lambda^4 }{16 \pi^2} \int d^4 x \sqrt{-g} \left[ - \frac{M_{pl}^2}{\Lambda^2} \frac{4 \pi^2}{\xi} u^2 + f_4
	+ \frac{R ~u^2}{6 \Lambda^2} f_1 \right. \\
	\nonumber {} & ~~~~~ + \frac{2}{\Lambda^4} \left( - \frac{1}{120} \nabla^2 R + \frac{1}{288} R^2
	- \frac{1}{180} R_{\mu \nu} R^{\mu \nu}
	- \frac{7}{1440} R_{\mu \nu \rho \sigma} R^{\mu \nu \rho \sigma} \right) \left( f_1 + \log \left( \frac{1}{y^2} \right) \right)  \\
	{} & ~~~~~ - \frac{2}{\Lambda^4} \left( \frac{1}{12} \nabla^2 R + \frac{1}{36} R_{\mu \nu} R^{\mu \nu}
	+ \frac{1}{144} R_{\mu \nu \rho \sigma} R^{\mu \nu \rho \sigma} \right) f_2  \\
	\nonumber {} & ~~~~~ \left. + \frac{2}{\Lambda^4} \left( \frac{1}{10} \nabla^2 R + \frac{1}{72} R^2
	+ \frac{7}{180} R_{\mu \nu} R^{\mu \nu} + \frac{1}{60} R_{\mu \nu \rho \sigma} R^{\mu \nu \rho \sigma} \right) f_3
	\right] ,
\end{align}
with:
\bb\label{def_f_1_2}
f_1 \equiv \frac{1}{1 + u^2} - \frac{y^2}{y^2 + u^2}
- \log \left( \frac{1 + u^2}{y^2 + u^2} \right) ;
\5\5
f_2 \equiv \frac{1}{2 ( 1 + u^2 )^2} - \frac{y^4}{2 ( y^2 + u^2 )^2} ;
\ee
\bb\label{def_f_3}
f_3 \equiv \frac{1}{3 ( 1 + u^2 )^3} - \frac{y^6}{3 ( y^2 + u^2 )^3} ;
\ee
\bb\label{def_f_4}
f_4 \equiv (1 - y^2) u^2
+ \log \left( 1 + u^2 \right)
- y^4 \log \left( 1 + \frac{u^2}{y^2} \right)
- u^4 \log \left( \frac{1 + u^2}{y^2 + u^2} \right) .
\ee
\bb\label{def_dimless_vars}
u \equiv \frac{\sigma}{\Lambda} ;
\5\5\5
y \equiv \frac{\epsilon}{\Lambda} .
\ee
The first line of (\ref{eff_action_CC}) was also obtained in \cite{Inagaki:1993ya} (with $y=0$), while the second, third, and fourth lines, i.e., the curvature${}^2$ corrections, are new.

%%%%%%%%%%%%%%%%%%%%%%%%%%%%%%%%%%%%%%%%%%%%%%%%%%%%%%%%%%%%%%%%%%%%%%%%%%

\subsection{\large Derivative Corrections to the NJL Effective Action}

%%%%%%%%%%%%%%%%%%%%%%%%%%%%%%%%%%%%%%%%%%%%%%%%%%%%%%%%%%%%%%%%%%%%%%%%%%

To capture the derivative corrections to the NJL effective action, the algorithm developed by \textit{M. K. Gaillard} \cite{Gaillard:1985uh} is used. Her metric is mostly minus, so the formulas below have been modified to fit our conventions. The goal is to calculate corrections up to the second order in derivatives, and since these derivatives act on a scalar, the curvature of the space-time will be ignored. The spinor action is once again the starting point:
\bb
\int d^4 x \overline{\Psi} \Big( i \slashed{\partial} - \sigma \Big) \Psi .
\ee
After integrating out the spinor field, the one-loop effective Lagrangian is:
\bb\label{eff_lag_kinetic}
\mathcal{L} \propto \int d^4 p ~ \text{tr} \log \Big( p^2 + \hat{M}^2 - i \widehat{\slashed{D} M} \Big) ,
\ee
where:
\bb\label{kin_corr_1}
\hat{M} = \sum_{n=0}^{+\infty} \frac{i^n}{n!} \Big( \partial_{\mu_1} \cdots \partial_{\mu_n} \sigma \Big)
\frac{\partial^n}{\partial p_{\mu_1} \cdots \partial p_{\mu_n}} =
\sigma + i \partial_{\mu} \sigma \frac{\partial}{\partial p_{\mu}}
- \frac{1}{2} \partial_{\mu} \partial_{\nu} \sigma \frac{\partial^2}{\partial p_{\mu} \partial p_{\nu}}
+ \mathcal{O} \Big( \partial^3 \sigma \Big) ;
\ee
\bb\label{kin_corr_2}
\widehat{\slashed{D} M} = \sum_{n=0}^{+\infty} \frac{i^n}{n!} \Big( \partial_{\mu_1} \cdots \partial_{\mu_n} \slashed{\partial} \sigma \Big)
\frac{\partial^n}{\partial p_{\mu_1} \cdots \partial p_{\mu_n}} =  \slashed{\partial} \sigma + i \partial_{\mu} \slashed{\partial} \sigma \frac{\partial}{\partial p_{\mu}}
+ \mathcal{O} \Big( \partial^3 \sigma \Big) .
\ee
Note that the presence of imaginary unit $i$ in front of the last term in (\ref{eff_lag_kinetic}) is essential; without it, the kinetic term for $\sigma$ one gets as a result of this procedure would have the wrong sign when $u = \sigma / \Lambda \ll 1$. This imaginary unit is missing in the original formula (4.21) in \cite{Gaillard:1985uh}. The reason is that while transitioning from formula (4.7) to (4.14), using (4.8) and (4.10-4.12), the imaginary unit is dropped, and then the typo/mistake propagates. After recovering it, one indeed arrives to (\ref{eff_lag_kinetic}). The presence of the imaginary unit can also be justified from the following reasoning: dropping all $\partial / \partial p$ corrections in (\ref{kin_corr_1}) and (\ref{kin_corr_2}) one has:
\bb
p^2 + \hat{M}^2 - i \widehat{\slashed{D} M} = p^2 + \sigma^2 - i \slashed{\partial} \sigma \rightarrow
- \partial^2 + \sigma^2 - i \slashed{\partial} \sigma ,
\ee
which does coincide with the ``square" of the Dirac operator:
\bb
\left( i \slashed{\partial} - \sigma \right) \left( i \slashed{\partial} - \sigma \right)^\dagger = - \left( i \slashed{\partial} - \sigma \right) \left( i \slashed{\partial} + \sigma \right) =   - \partial^2 + \sigma^2 - i \slashed{\partial} \sigma .
\ee
The integrand in (\ref{eff_lag_kinetic}) has to be understood in the following way:
\begin{align*}
	{} & \resizebox{1\hsize}{!}{ $ \log \Big( p^2 + \hat{M}^2 - i \widehat{\slashed{D} M} \Big)
		= \log \left( p^2 + \sigma^2 - \frac12 \partial_\mu \partial_\nu \sigma^2 \frac{\partial^2}{\partial p_{\mu} \partial p_{\nu}} + 2 i \sigma \partial_\mu \sigma \frac{\partial}{\partial p_{\mu}} - i \slashed{\partial} \sigma
		+ \partial_\mu \slashed{\partial} \sigma \frac{\partial}{\partial p_{\mu}} \right) $ } \\
	{} & \resizebox{1\hsize}{!}{ $ = \log \left( p^2 + \sigma^2 \right) + \log \left[ 1 +
		\left( - \frac12 \partial_\mu \partial_\nu \sigma^2 \frac{\partial^2}{\partial p_{\mu} \partial p_{\nu}} + 2 i \sigma \partial_\mu \sigma \frac{\partial}{\partial p_{\mu}} - i \slashed{\partial} \sigma
		+ \partial_\mu \slashed{\partial} \sigma \frac{\partial}{\partial p_{\mu}} \right) \frac{1}{p^2 + \sigma^2} \right]
		$. }
\end{align*}
The second logarithm has to be Taylor expanded up to the second order in $\partial \sigma$. After performing symmetrization of the momentum integral, Wick rotating the $p_0$ and imposing IR and UV cut-offs:
\bb\label{eff_lag_kinetic_fin}
\frac{1}{\Lambda^4} \mathcal{L} \propto ~ f_4 + \frac{1}{\Lambda^2} \left( \partial u \right)^2 f_0 ,
\ee
with:
\bb\label{def_f_0}
f_0 \equiv f_1 + 4 f_3 +
8 u^2 \left( \frac{1}{(1+u^2)^3} - \frac{y^4}{(y^2+u^2)^3}  \right) .
\ee
In (\ref{eff_lag_kinetic_fin}), the first term is the usual loop contribution, while the second term is the kinetic correction to it. We are after the latter since the first term is already present in (\ref{eff_action_CC}). The next order corrections to (\ref{eff_lag_kinetic_fin}) would involve terms like $\left( \partial \sigma \right)^4 $, $\partial^2 \sigma \left( \partial \sigma \right)^2 $ and $\partial \sigma \partial \sigma \partial \partial \sigma$, as well as $R \left( \partial \sigma \right)^2$ and $R_{\mu \nu} \partial^\mu \sigma \partial^\nu \sigma$ when the curvature of the space-time is taken into account. These terms will be ignored in the current analysis.

%%%%%%%%%%%%%%%%%%%%%%%%%%%%%%%%%%%%%%%%%%%%%%%%%%%%%%%%%%%%%%%%%%%%%%%%%%

\subsection{\large Comparison With the Heat Kernel Results}

%%%%%%%%%%%%%%%%%%%%%%%%%%%%%%%%%%%%%%%%%%%%%%%%%%%%%%%%%%%%%%%%%%%%%%%%%%

Combing the results of the two previous subsections, one arrives at the final NJL action involving both the kinetic and the curvature corrections:
\begin{align}\label{eff_action_CC_kin}
	\nonumber {} & S_{eff} = \frac{ \Lambda^4 }{16 \pi^2} \int d^4 x \sqrt{-g} \left[ - \frac{M_{pl}^2}{\Lambda^2} \frac{4 \pi^2}{\xi} u^2 + f_4 + \frac{1}{\Lambda^2} \left( \partial u \right)^2 f_0
	+ \frac{R ~u^2}{6 \Lambda^2} f_1  \right. \\
	\nonumber {} & ~~~~~ + \frac{2}{\Lambda^4} \left( - \frac{1}{120} \nabla^2 R + \frac{1}{288} R^2
	- \frac{1}{180} R_{\mu \nu} R^{\mu \nu}
	- \frac{7}{1440} R_{\mu \nu \rho \sigma} R^{\mu \nu \rho \sigma} \right) \left( f_1 + \log \left( \frac{1}{y^2} \right) \right)  \\
	{} & ~~~~~ - \frac{2}{\Lambda^4} \left( \frac{1}{12} \nabla^2 R + \frac{1}{36} R_{\mu \nu} R^{\mu \nu}
	+ \frac{1}{144} R_{\mu \nu \rho \sigma} R^{\mu \nu \rho \sigma} \right) f_2  \\
	\nonumber {} & ~~~~~ \left. + \frac{2}{\Lambda^4} \left( \frac{1}{10} \nabla^2 R + \frac{1}{72} R^2
	+ \frac{7}{180} R_{\mu \nu} R^{\mu \nu} + \frac{1}{60} R_{\mu \nu \rho \sigma} R^{\mu \nu \rho \sigma} \right) f_3
	\right] .
\end{align}
Setting the IR scale $y = 0$, and going in the limit $\Lambda \rightarrow + \infty$, the action (\ref{eff_action_CC_kin}) simplifies significantly:
\begin{align}\label{eff_action_CC_kin_small}
	\nonumber {} & S_{eff} = \frac{ 1 }{16 \pi^2} \int d^4 x \sqrt{-g} \left[ - M_{pl}^2 \frac{4 \pi^2}{\xi} \sigma^2
	+ 2 \hbar \sigma^2 \Lambda^2 - 2 \hbar \left( \frac12 \sigma^4 + \frac12 \left( \partial \sigma \right)^2
	+ \frac{R \sigma^2}{12} \right.  \right. \\
	{} & ~~~~~  \left. \left. - \frac{1}{120} \nabla^2 R + \frac{1}{288} R^2
	- \frac{1}{180} R_{\mu \nu} R^{\mu \nu}
	- \frac{7}{1440} R_{\mu \nu \rho \sigma} R^{\mu \nu \rho \sigma} \right) \log \left( \frac{\Lambda^2}{\sigma^2} \right) \right] .
\end{align}
In the last formula, $\hbar$ was recovered to highlight the quantum nature of the relevant terms, and only those were kept that diverge as $\Lambda \rightarrow + \infty$. \\

Terms proportional to $\hbar$ can also be calculated using the Heat Kernel method. Let's follow the papers by \textit{Barvinsky et al.} \cite{Barvinsky:1990up,Barvinsky:1993en}. The loop contribution can be written as the logarithm from the functional determinant of the Dirac operator, i.e.:
\bb\label{loop_det}
\log \det \Big( g^{\mu \nu} \nabla_\mu \nabla_\nu - \frac{1}{4} R + i \gamma^\mu \left( \nabla_\mu \sigma \right)  - \sigma^2 \Big) .
\ee
According to \cite{Barvinsky:1990up,Barvinsky:1993en}:
\bb\label{loop_det_heat_ker}
\log \det \Big( g^{\mu \nu} \nabla_\mu \nabla_\nu + \hat{P} - \frac{1}{6} R \Big) = - \int \frac{ d \tau }{\tau} K (\tau) ,
\ee
where $K(\tau)$ is the Heat Kernel and $\tau$ is the proper time. The lower limit of $\tau$ corresponds to the UV limit in momentum. In the limit $\tau \rightarrow 0$ (Schwinger-DeWitt), the Heat Kernel can be Taylor expanded in powers of $\tau$:
\begin{align}\label{orig_HK}
	\nonumber K (\tau) = & \frac{1}{(4 \pi)^2 \tau^2} \int d^4 x \sqrt{-g} ~\text{tr} \left[ 1 + \tau \hat{P} + \tau^2
	\left( \frac12 \hat{P}^2 - \frac{1}{120} \nabla^2 R
	+ \frac{1}{768} \mathcal{R}_{\mu \nu} \mathcal{R}^{\mu \nu}  \right. \right. \\
	{} & ~~~~~~~~~~~~~~~~~~~~~~~~~~~~~~ \left. \left. - \frac{1}{180} R_{\mu \nu} R^{\mu \nu} + \frac{1}{180} R_{\mu \nu \rho \sigma} R^{\mu \nu \rho \sigma} \right) \right] .
\end{align}
Note that in \cite{Barvinsky:1990up,Barvinsky:1993en}, the term $\nabla^2 R$ is neglected as it's a total derivative. This term can be recovered by consulting \cite{DeWitt:1965jb}. Comparing (\ref{loop_det}) with (\ref{loop_det_heat_ker}) yields:
\bb
\hat{P} \equiv - \frac{1}{12} R - \sigma^2 + i \gamma^\mu \nabla_\mu \sigma .
\ee
After plugging this into (\ref{orig_HK}):
\begin{align}\label{HK_mine}
	\nonumber K (\tau) = & \frac{1}{(4 \pi)^2 \tau^2} \int d^4 x \sqrt{g} ~ \left[ 1 - \tau \left( \sigma^2 + \frac{1}{12} R \right) + \tau^2 \left( \frac12 \sigma^4 + \frac12 (\partial \sigma )^2 + \frac{1}{12} \sigma^2 R \right. \right. \\
	{} & ~~~~~~~~~ \left. \left. - \frac{1}{120} \nabla^2 R
	+ \frac{1}{288} R^2
	- \frac{1}{180} R_{\mu \nu} R^{\mu \nu} - \frac{7}{1440} R_{\mu \nu \rho \sigma} R^{\mu \nu \rho \sigma} \right) \right]
	\text{tr} \mathds{1} .
\end{align}
In the Heat Kernel regularization, the loop integrals are finite, and the UV divergences manifest themselves in the lower limit of the proper time integrals. Cutting the proper time integral at $\tau = \Lambda^{-2}$ in the lower limit and at $\tau = \sigma^{-2}$ in the upper limit and keeping only the terms that diverge as $\Lambda \rightarrow + \infty$ one finally obtains:
\begin{align*}
	{} & \log \det \Big( g^{\mu \nu} \nabla_\mu \nabla_\nu - \frac{1}{4} R + i \gamma^\mu \left( \nabla_\mu \sigma \right)  - \sigma^2 \Big) = \\
	{} & \frac{1}{(4 \pi)^2 } \int d^4 x \sqrt{-g} ~ \Bigg[ \Lambda^2 \sigma^2 -
	\left( \frac12 \sigma^4  + \frac12 (\partial \sigma )^2 + \frac{1}{12} \sigma^2 R - \frac{1}{120} \nabla^2 R \right. \\
	{} & ~~~~~~~~~~~~~~~~~~~~~~~~ \left. + \frac{1}{288} R^2
	- \frac{1}{180} R_{\mu \nu} R^{\mu \nu} - \frac{7}{1440} R_{\mu \nu \rho \sigma} R^{\mu \nu \rho \sigma} \right)
	\log \left( \frac{\Lambda^2}{\sigma^2} \right) \Bigg]
	\text{tr} \mathds{1} .
\end{align*}
This formula is identical to the $\hbar$ part of (\ref{eff_action_CC_kin_small}). \\

Given the function $F(p)$, one can guess that the next order curvature corrections to it, symbolically, must have the following form:
\bb
F(p) \sim \frac{R^n}{10^n (p^2+s^2)^{(n+1)}} + \cdots .
\ee
The ``of order" factor $~10^{-n}$ can be observed from formulas (4.47)-(4.49) in \cite{Barvinsky:1993en}. At the level of effective action, these corrections translate as:
\bb
\Delta S_{eff} \sim \Lambda^4 \int d^4 x \sqrt{-g} ~ \frac{1}{10^n} ~\frac{R^n}{\Lambda^{2n}} ~q_n \left( \frac{\sigma^2}{\Lambda^2} \right) .
\ee
For $n>2$, the functions $q_n$ are polynomials of their argument when this argument is small; and satisfy $q_n (0) = 0$. The main difference with the case we studied in detail is that the linear and quadratic curvature corrections are additionally multiplied by the $\log (\sigma^2 / \Lambda^2)$, so when $\sigma \ll \Lambda ~(u \ll 1)$ they are less suppressed. 

From the analysis above, it follows that several hierarchies take place when expanding the effective action as a series in curvature. When $u \ll 1$, the smallness of $R/\Lambda^2$ is not necessary for the curvature truncation we made in (\ref{eff_action_CC}) to work. The latter is a feature of the four-dimensional space-time. For the solutions presented in section \ref{numerics}, both conditions: $R/\Lambda^2 \ll 1$ and $u \ll 1$, are satisfied, so ignoring terms of a higher degree in curvature is perfectly justified.

%%%%%%%%%%%%%%%%%%%%%%%%%%%%%%%%%%%%%%%%%%%%%%%%%%%%%%%%%%%%%%%%%%%%%%%%%%
%%%%%%%%%%%%%%%%%%%%%%%%%%%%%%%%%%%%%%%%%%%%%%%%%%%%%%%%%%%%%%%%%%%%%%%%%%
%%%%%%%%%%%%%%%%%%%%%%%%%%%%%%%%%%%%%%%%%%%%%%%%%%%%%%%%%%%%%%%%%%%%%%%%%%

\bibliographystyle{utphys}
\bibliography{refs}

\providecommand{\href}[2]{#2}\begingroup\raggedright\begin{thebibliography}{10}

\bibitem{Tukhashvili:2023itb}
G.~Tukhashvili and P.~J. Steinhardt, ``{Cosmological Bounces Induced by a
  Fermion Condensate},''
  \href{http://dx.doi.org/10.1103/PhysRevLett.131.091001}{{\em Phys. Rev.
  Lett.} {\bfseries 131} no.~9, (2023) 091001},
  \href{http://arxiv.org/abs/2307.16098}{{\ttfamily arXiv:2307.16098 [gr-qc]}}.

\bibitem{Cook:2020oaj}
W.~G. Cook, I.~A. Glushchenko, A.~Ijjas, F.~Pretorius, and P.~J. Steinhardt,
  ``{Supersmoothing through Slow Contraction},''
  \href{http://dx.doi.org/10.1016/j.physletb.2020.135690}{{\em Phys. Lett. B}
  {\bfseries 808} (2020) 135690},
  \href{http://arxiv.org/abs/2006.01172}{{\ttfamily arXiv:2006.01172 [gr-qc]}}.

\bibitem{Ijjas:2020dws}
A.~Ijjas, W.~G. Cook, F.~Pretorius, P.~J. Steinhardt, and E.~Y. Davies,
  ``{Robustness of slow contraction to cosmic initial conditions},''
  \href{http://dx.doi.org/10.1088/1475-7516/2020/08/030}{{\em JCAP} {\bfseries
  08} (2020) 030}, \href{http://arxiv.org/abs/2006.04999}{{\ttfamily
  arXiv:2006.04999 [gr-qc]}}.

\bibitem{Ijjas:2021gkf}
A.~Ijjas, A.~P. Sullivan, F.~Pretorius, P.~J. Steinhardt, and W.~G. Cook,
  ``{Ultralocality and slow contraction},''
  \href{http://dx.doi.org/10.1088/1475-7516/2021/06/013}{{\em JCAP} {\bfseries
  06} (2021) 013}, \href{http://arxiv.org/abs/2103.00584}{{\ttfamily
  arXiv:2103.00584 [gr-qc]}}.

\bibitem{Ijjas:2021wml}
A.~Ijjas, F.~Pretorius, P.~J. Steinhardt, and A.~P. Sullivan, ``{The effects of
  multiple modes and reduced symmetry on the rapidity and robustness of slow
  contraction},'' \href{http://dx.doi.org/10.1016/j.physletb.2021.136490}{{\em
  Phys. Lett. B} {\bfseries 820} (2021) 136490},
  \href{http://arxiv.org/abs/2104.12293}{{\ttfamily arXiv:2104.12293 [gr-qc]}}.

\bibitem{Kist:2022mew}
T.~Kist and A.~Ijjas, ``{The robustness of slow contraction and the shape of
  the scalar field potential},''
  \href{http://arxiv.org/abs/2205.01519}{{\ttfamily arXiv:2205.01519 [gr-qc]}}.

\bibitem{Penrose:1964wq}
R.~Penrose, ``{Gravitational collapse and space-time singularities},''
  \href{http://dx.doi.org/10.1103/PhysRevLett.14.57}{{\em Phys. Rev. Lett.}
  {\bfseries 14} (1965) 57--59}.

\bibitem{Hawking:1966jv}
S.~Hawking, ``{The Occurrence of singularities in cosmology. II},''
  \href{http://dx.doi.org/10.1098/rspa.1966.0255}{{\em Proc. Roy. Soc. Lond. A}
  {\bfseries 295} (1966) 490--493}.

\bibitem{Hawking:1970zqf}
S.~W. Hawking and R.~Penrose, ``{The Singularities of gravitational collapse
  and cosmology},'' \href{http://dx.doi.org/10.1098/rspa.1970.0021}{{\em Proc.
  Roy. Soc. Lond. A} {\bfseries 314} (1970) 529--548}.

\bibitem{Ijjas:2016tpn}
A.~Ijjas and P.~J. Steinhardt, ``{Classically stable nonsingular cosmological
  bounces},'' \href{http://dx.doi.org/10.1103/PhysRevLett.117.121304}{{\em
  Phys. Rev. Lett.} {\bfseries 117} no.~12, (2016) 121304},
  \href{http://arxiv.org/abs/1606.08880}{{\ttfamily arXiv:1606.08880 [gr-qc]}}.

\bibitem{Ijjas:2016vtq}
A.~Ijjas and P.~J. Steinhardt, ``{Fully stable cosmological solutions with a
  non-singular classical bounce},''
  \href{http://dx.doi.org/10.1016/j.physletb.2016.11.047}{{\em Phys. Lett. B}
  {\bfseries 764} (2017) 289--294},
  \href{http://arxiv.org/abs/1609.01253}{{\ttfamily arXiv:1609.01253 [gr-qc]}}.

\bibitem{Easson:2011zy}
D.~A. Easson, I.~Sawicki, and A.~Vikman, ``{G-Bounce},''
  \href{http://dx.doi.org/10.1088/1475-7516/2011/11/021}{{\em JCAP} {\bfseries
  11} (2011) 021}, \href{http://arxiv.org/abs/1109.1047}{{\ttfamily
  arXiv:1109.1047 [hep-th]}}.

\bibitem{Cai:2012va}
Y.-F. Cai, D.~A. Easson, and R.~Brandenberger, ``{Towards a Nonsingular
  Bouncing Cosmology},''
  \href{http://dx.doi.org/10.1088/1475-7516/2012/08/020}{{\em JCAP} {\bfseries
  08} (2012) 020}, \href{http://arxiv.org/abs/1206.2382}{{\ttfamily
  arXiv:1206.2382 [hep-th]}}.

\bibitem{Cai:2013vm}
Y.-F. Cai, R.~Brandenberger, and P.~Peter, ``{Anisotropy in a Nonsingular
  Bounce},'' \href{http://dx.doi.org/10.1088/0264-9381/30/7/075019}{{\em Class.
  Quant. Grav.} {\bfseries 30} (2013) 075019},
  \href{http://arxiv.org/abs/1301.4703}{{\ttfamily arXiv:1301.4703 [gr-qc]}}.

\bibitem{Libanov:2016kfc}
M.~Libanov, S.~Mironov, and V.~Rubakov, ``{Generalized Galileons: instabilities
  of bouncing and Genesis cosmologies and modified Genesis},''
  \href{http://dx.doi.org/10.1088/1475-7516/2016/08/037}{{\em JCAP} {\bfseries
  08} (2016) 037}, \href{http://arxiv.org/abs/1605.05992}{{\ttfamily
  arXiv:1605.05992 [hep-th]}}.

\bibitem{Trautman:1973wy}
A.~Trautman, ``{Spin and torsion may avert gravitational singularities},''
  \href{http://dx.doi.org/10.1038/physci242007a0}{{\em Nature} {\bfseries 242}
  (1973) 7--8}.

\bibitem{Isham:1974ci}
C.~J. Isham and J.~E. Nelson, ``{Quantization of a Coupled Fermi Field and
  Robertson-Walker Metric},''
  \href{http://dx.doi.org/10.1103/PhysRevD.10.3226}{{\em Phys. Rev. D}
  {\bfseries 10} (1974) 3226}.

\bibitem{Alexander:2008vt}
S.~Alexander and T.~Biswas, ``{The Cosmological BCS mechanism and the Big Bang
  Singularity},'' \href{http://dx.doi.org/10.1103/PhysRevD.80.023501}{{\em
  Phys. Rev. D} {\bfseries 80} (2009) 023501},
  \href{http://arxiv.org/abs/0807.4468}{{\ttfamily arXiv:0807.4468 [hep-th]}}.

\bibitem{Magueijo:2012ug}
J.~a. Magueijo, T.~G. Zlosnik, and T.~W.~B. Kibble, ``{Cosmology with a
  spin},'' \href{http://dx.doi.org/10.1103/PhysRevD.87.063504}{{\em Phys. Rev.
  D} {\bfseries 87} no.~6, (2013) 063504},
  \href{http://arxiv.org/abs/1212.0585}{{\ttfamily arXiv:1212.0585
  [astro-ph.CO]}}.

\bibitem{Sciama:1958we}
D.~Sciama, ``{On a Nonsymmetric theory of the pure gravitational field},''
  \href{http://dx.doi.org/10.1017/S030500410003320X}{{\em Proc. Cambridge Phil.
  Soc.} {\bfseries 54} (1958) 72}.

\bibitem{Kibble:1961ba}
T.~W.~B. Kibble, ``{Lorentz invariance and the gravitational field},''
  \href{http://dx.doi.org/10.1063/1.1703702}{{\em J. Math. Phys.} {\bfseries 2}
  (1961) 212--221}.

\bibitem{Hehl:1976kj}
F.~W. Hehl, P.~Von Der~Heyde, G.~D. Kerlick, and J.~M. Nester, ``{General
  Relativity with Spin and Torsion: Foundations and Prospects},''
  \href{http://dx.doi.org/10.1103/RevModPhys.48.393}{{\em Rev. Mod. Phys.}
  {\bfseries 48} (1976) 393--416}.

\bibitem{Shapiro:2001rz}
I.~L. Shapiro, ``{Physical aspects of the space-time torsion},''
  \href{http://dx.doi.org/10.1016/S0370-1573(01)00030-8}{{\em Phys. Rept.}
  {\bfseries 357} (2002) 113},
  \href{http://arxiv.org/abs/hep-th/0103093}{{\ttfamily arXiv:hep-th/0103093}}.

\bibitem{Weyl:1950xa}
H.~Weyl, ``{A Remark on the coupling of gravitation and electron},''
  \href{http://dx.doi.org/10.1103/PhysRev.77.699}{{\em Phys. Rev.} {\bfseries
  77} (1950) 699--701}.

\bibitem{Nambu:1961tp}
Y.~Nambu and G.~Jona-Lasinio, ``{Dynamical Model of Elementary Particles Based
  on an Analogy with Superconductivity. 1.},''
  \href{http://dx.doi.org/10.1103/PhysRev.122.345}{{\em Phys. Rev.} {\bfseries
  122} (1961) 345--358}.

\bibitem{Farnsworth:2017wzr}
S.~Farnsworth, J.-L. Lehners, and T.~Qiu, ``{Spinor driven cosmic bounces and
  their cosmological perturbations},''
  \href{http://dx.doi.org/10.1103/PhysRevD.96.083530}{{\em Phys. Rev. D}
  {\bfseries 96} no.~8, (2017) 083530},
  \href{http://arxiv.org/abs/1709.03171}{{\ttfamily arXiv:1709.03171 [gr-qc]}}.

\bibitem{Inagaki:1993ya}
T.~Inagaki, T.~Muta, and S.~D. Odintsov, ``{Nambu-Jona-Lasinio model in curved
  space-time},'' \href{http://dx.doi.org/10.1142/S0217732393001835}{{\em Mod.
  Phys. Lett. A} {\bfseries 8} (1993) 2117--2124},
  \href{http://arxiv.org/abs/hep-th/9306023}{{\ttfamily arXiv:hep-th/9306023}}.

\bibitem{Misner:1973prb}
C.~W. Misner, K.~S. Thorne, and J.~A. Wheeler, {\em {Gravitation}}.
\newblock 1973.

\bibitem{Ijjas:2021ewd}
A.~Ijjas and R.~Kolevatov, ``{Nearly scale-invariant curvature modes from
  entropy perturbations during the graceful exit phase},''
  \href{http://dx.doi.org/10.1103/PhysRevD.103.L101302}{{\em Phys. Rev. D}
  {\bfseries 103} no.~10, (2021) L101302},
  \href{http://arxiv.org/abs/2102.03818}{{\ttfamily arXiv:2102.03818 [gr-qc]}}.

\bibitem{Ijjas:2020cyh}
A.~Ijjas and R.~Kolevatov, ``{Sourcing curvature modes with entropy
  perturbations in non-singular bouncing cosmologies},''
  \href{http://dx.doi.org/10.1088/1475-7516/2021/06/012}{{\em JCAP} {\bfseries
  06} (2021) 012}, \href{http://arxiv.org/abs/2012.08249}{{\ttfamily
  arXiv:2012.08249 [gr-qc]}}.

\bibitem{Bagrov:1994re}
V.~G. Bagrov, I.~L. Buchbinder, and I.~L. Shapiro, ``{On the possible
  experimental manifestations of the torsion field at low-energies},''
  \href{http://arxiv.org/abs/hep-th/9406122}{{\ttfamily arXiv:hep-th/9406122}}.

\bibitem{Hammond:1995rp}
R.~T. Hammond, ``{Upper limit on the torsion coupling constant},''
  \href{http://dx.doi.org/10.1103/PhysRevD.52.6918}{{\em Phys. Rev. D}
  {\bfseries 52} (1995) 6918--6921}.

\bibitem{Hammond:1996ua}
R.~T. Hammond, ``{Helicity flip cross-section from gravity with torsion},''
  \href{http://dx.doi.org/10.1088/0264-9381/13/7/002}{{\em Class. Quant. Grav.}
  {\bfseries 13} (1996) 1691--1697}.

\bibitem{Chui:1993jp}
T.~C.~P. Chui and W.-T. Ni, ``{Experimental search for an anomalous spin spin
  interaction between electrons},''
  \href{http://dx.doi.org/10.1103/PhysRevLett.71.3247}{{\em Phys. Rev. Lett.}
  {\bfseries 71} (1993) 3247--3250}.

\bibitem{Shifman:1978bx}
M.~A. Shifman, A.~I. Vainshtein, and V.~I. Zakharov, ``{QCD and Resonance
  Physics. Theoretical Foundations},''
  \href{http://dx.doi.org/10.1016/0550-3213(79)90022-1}{{\em Nucl. Phys. B}
  {\bfseries 147} (1979) 385--447}.

\bibitem{Shapiro:1994vs}
I.~L. Shapiro, ``{The Interaction of quantized matter fields with torsion and
  metric: NJL model},'' \href{http://dx.doi.org/10.1142/S021773239400054X}{{\em
  Mod. Phys. Lett. A} {\bfseries 9} (1994) 729--733}.

\bibitem{Ford:1981xj}
L.~H. Ford and D.~J. Toms, ``{Dynamical Symmetry Breaking Due to Radiative
  Corrections in Cosmology},''
  \href{http://dx.doi.org/10.1103/PhysRevD.25.1510}{{\em Phys. Rev. D}
  {\bfseries 25} (1982) 1510}.

\bibitem{Alexander:2014uaa}
S.~Alexander, Y.-F. Cai, and A.~Marciano, ``{Fermi-bounce cosmology and the
  fermion curvaton mechanism},''
  \href{http://dx.doi.org/10.1016/j.physletb.2015.04.026}{{\em Phys. Lett. B}
  {\bfseries 745} (2015) 97--104},
  \href{http://arxiv.org/abs/1406.1456}{{\ttfamily arXiv:1406.1456 [gr-qc]}}.

\bibitem{Erickson:2003zm}
J.~K. Erickson, D.~H. Wesley, P.~J. Steinhardt, and N.~Turok, ``{Kasner and
  mixmaster behavior in universes with equation of state w \ensuremath{>}=
  1},'' \href{http://dx.doi.org/10.1103/PhysRevD.69.063514}{{\em Phys. Rev. D}
  {\bfseries 69} (2004) 063514},
  \href{http://arxiv.org/abs/hep-th/0312009}{{\ttfamily arXiv:hep-th/0312009}}.

\bibitem{Weinberg:2008zzc}
S.~Weinberg, {\em {Cosmology}}.
\newblock 2008.

\bibitem{Pirani:1956wr}
F.~A.~E. Pirani, ``{Invariant formulation of gravitational radiation theory},''
  \href{http://dx.doi.org/10.1103/PhysRev.105.1089}{{\em Phys. Rev.} {\bfseries
  105} (1957) 1089--1099}.

\bibitem{Petrov:1969}
A.~Z. {Petrov}, {\em {Einstein Spaces}}.
\newblock 1969.

\bibitem{Brewin:1996yk}
L.~Brewin, ``{Riemann normal coordinates, smooth lattices and numerical
  relativity},'' \href{http://dx.doi.org/10.1088/0264-9381/15/10/014}{{\em
  Class. Quant. Grav.} {\bfseries 15} (1998) 3085--3120},
  \href{http://arxiv.org/abs/gr-qc/9701057}{{\ttfamily arXiv:gr-qc/9701057}}.

\bibitem{Bunch:1979uk}
T.~S. Bunch and L.~Parker, ``{Feynman Propagator in Curved Space-Time: A
  Momentum Space Representation},''
  \href{http://dx.doi.org/10.1103/PhysRevD.20.2499}{{\em Phys. Rev. D}
  {\bfseries 20} (1979) 2499--2510}.

\bibitem{Parker:1983pe}
L.~Parker and D.~J. Toms, ``{Renormalization Group Analysis of Grand Unified
  Theories in Curved Space-time},''
  \href{http://dx.doi.org/10.1103/PhysRevD.29.1584}{{\em Phys. Rev. D}
  {\bfseries 29} (1984) 1584}.

\bibitem{Gaillard:1985uh}
M.~K. Gaillard, ``{The Effective One Loop Lagrangian With Derivative
  Couplings},'' \href{http://dx.doi.org/10.1016/0550-3213(86)90264-6}{{\em
  Nucl. Phys. B} {\bfseries 268} (1986) 669--692}.

\bibitem{Barvinsky:1990up}
A.~O. Barvinsky and G.~A. Vilkovisky, ``{Covariant perturbation theory. 2:
  Second order in the curvature. General algorithms},''
  \href{http://dx.doi.org/10.1016/0550-3213(90)90047-H}{{\em Nucl. Phys. B}
  {\bfseries 333} (1990) 471--511}.

\bibitem{Barvinsky:1993en}
A.~O. Barvinsky, Y.~V. Gusev, V.~V. Zhytnikov, and G.~A. Vilkovisky,
  ``{Covariant perturbation theory. 4. Third order in the curvature},''
  \href{http://arxiv.org/abs/0911.1168}{{\ttfamily arXiv:0911.1168 [hep-th]}}.

\bibitem{DeWitt:1965jb}
B.~S. DeWitt, ``{Dynamical theory of groups and fields},'' {\em Conf. Proc.}
  {\bfseries C630701} (1964) 585--820.
[Les Houches Lect. Notes13,585(1964)].
%%CITATION = CONFP,C630701,585;%%.

\end{thebibliography}\endgroup

%%%%%%%%%%%%%%%%%%%%%%%%%%%%%%%%%%%%%%%%%%%%%%%%%%%%%%%%%%%%%%%%%%%%%
%%%%%%%%%%%%%%%%%%%%%%%%%%%%%%%%%%%%%%%%%%%%%%%%%%%%%%%%%%%%%%%%%%%%%
%%%%%%%%%%%%%%%%%%%%%%%%%%%%%%%%%%%%%%%%%%%%%%%%%%%%%%%%%%%%%%%%%%%%%

\end{document}